\documentclass[12pt, draftclsnofoot,onecolumn]{IEEEtran}
\IEEEoverridecommandlockouts
\usepackage[english]{babel}
\usepackage{blindtext}
\usepackage{algorithm}
\usepackage{algorithmic}
\usepackage{graphicx}
\usepackage{subfigure}
\usepackage{amsmath}
\usepackage{amssymb}
\usepackage{cases}
\usepackage{setspace}
\usepackage{multirow}
\usepackage{mathrsfs}

\begin{document}
\title{Efficient Scheduling and Power Allocation for D2D-assisted Wireless Caching Networks}
\author{\authorblockN{Lin Zhang, \IEEEmembership{Student Member,~IEEE}, Ming Xiao, \IEEEmembership{Senior Member,~IEEE},\\ Gang Wu, \IEEEmembership{Member,~IEEE}, and Shaoqian Li, \IEEEmembership{Fellow,~IEEE}}
\thanks{Lin Zhang, Gang Wu, and Shaoqian Li are with the National Key Lab of Science and Technology on Communications, University of Electronic Science and Technology of China, Chengdu, China, emails: linzhang1913@gmail.com, \{wugang99, lsq\}@uestc.edu.cn; Ming Xiao is with the School of Electrical Engineering, Royal Institute of Technology, KTH, email: mxiao@ieee.org.}
}\maketitle

\thispagestyle{empty}

\begin{abstract}
We study an one-hop device-to-device (D2D) assisted wireless caching
network, where popular files are randomly and independently cached in the memory of
end-users. Each user may obtain the requested files from
its own memory without any transmission, or from a helper through an
one-hop D2D transmission, or from the base station (BS). We formulate a joint D2D link scheduling and power allocation problem to maximize the system throughput. However, the problem is non-convex and obtaining an optimal solution is computationally hard. Alternatively, we decompose the problem into a D2D link scheduling problem and an optimal power allocation problem. To solve the two subproblems, we first develop a D2D link scheduling algorithm to select the largest number of D2D links satisfying both the signal to interference plus noise ratio (SINR) and the transmit power constraints. Then, we develop an optimal power allocation algorithm to maximize the minimum transmission rate of the scheduled D2D links. Numerical results indicate that both the number of the scheduled D2D links and the system throughput can be improved simultaneously with the Zipf-distribution caching scheme, the proposed D2D link scheduling algorithm, and the proposed optimal power allocation algorithm compared with the state of arts.
\end{abstract}


\newpage

\section{Introduction}
With the emergence of various applications, especially multimedia, mobile data has been explosively increasing in recent years. It is predicted that the mobile traffic will grow by more than 200 to 2000 times in the next
few years according to a study from IMT 2020 \cite{IMT}. The huge data amount pushes operators to provide high-throughput wireless access services.
However, the current wireless access technology is already quite close to the performance limits and thus new communication strategies are needed to meet the increasing requirement from mobile subscribers.

%, i.e., 5G, since 5G is supposed to
%increase the spectral efficiency by 5 to 15 times compared to
%existing 4G cellular network and the supported data traffic in 5G
%can at most increase by 15 times if no more spectrum can be used compared with that in 4G. This
%means the data traffic can hardly be satisfied in the next few
%years. Thus, it is crucial to develop an alternative to cope with
%the highly increasing data traffic demand.

One promising approach is to use heterogeneous networks
\cite{heter1}, \cite{heter2}, where one cell is divided into multiple§
small cells, i.e., microcells, picocells, and femtocells. Within
each small cell, one \emph{low-power base station} (LPBS) is
equipped to serve the users in coverage. Specially, each LPBS is
connected to the \emph{base station} (BS) of the cell with a
backhaul, e.g., high-speed fiber \cite{survey_fiber}.
Then, the requested files by a user are first transmitted from the BS to the
LPBS through the backhaul and then transmitted from the
LPBS to the user. The deployment of LPBS shortens the wireless transmission distance and
results in a spatial gain, which boosts the system throughput
dramatically compared with the conventional cellular networks.
However, it may be very expensive to establish and maintain the LPBS
as well as the backhaul. This hinders the application of the heterogeneous
networks in practice.

Alternatively, wireless caching recently has attracted a lot of
attentions for the advantages of fast response and without
heavily relying on the backhaul \cite{survey1}, \cite{survey2}. As observed that some files, e.g., video clips, usually remain
popular in a certain period of time, say one day, one week, even one
month. Meanwhile, with the fast development of \emph{integrated circuit} (IC) technologies, the
price of storage memory drops quickly. Then, the storage capacity in devices can be utilized for wireless caching. Specifically, some popular
files can be cached into the memory of users during the off-load time, say
middle night, such that users may obtain the requested files from
the devices rather than the BS in the peak-load time, say daytime.
Notably, a popular file may be requested by multiple
users. Thus, more user requests can be accommodated with the help of wireless caching during the peak-load time. This increases the system throughput significantly. Besides, high-speed
backhauls are not required for the wireless caching since the cached
files can be transmitted through wireless channels during the off-load time.

In this paper, we focus on the wireless caching networks with one-hop
\emph{device-to-device} (D2D) communication among users, since D2D
communication has been shown to be a promising candidate to improve
the system throughput \cite{D2Dsurvey}. Then, some popular files are
cached into the memory of users and each user may obtain the
requested files from its own memory without any transmission,
or from a helper through an one-hop D2D transmission, or from the BS. We formulate a joint D2D link scheduling and power allocation problem to maximize the system throughput. However, the problem is non-convex and obtaining an optimal solution is computationally hard. Alternatively, we seek to obtain a suboptimal solution with reasonable complexity. Briefly, we intend to schedule the D2D links with strong communication channels and weak interference channels and allocate the power to the scheduled D2D links fairly. Thus, we decompose the joint optimization problem into a D2D link scheduling problem and an optimal power allocation problem. To solve the two subproblems, we first develop a D2D link scheduling algorithm to select the largest number of D2D links satisfying both the \emph{signal to interference plus noise ratio} (SINR) and the transmit power constraints. Then, we develop an optimal power allocation algorithm to maximize the minimum transmission rate of the scheduled D2D links. Numerical results indicate that both the number of the scheduled D2D links and the system throughput can be improved simultaneously with the Zipf-distribution caching scheme, the proposed D2D link scheduling algorithm, and the proposed optimal power allocation algorithm compared with the state of arts.

\emph{Related literature:} Wireless caching has recently been studied from
theoretical perspectives in \cite{Mod-Ali-14}-\cite{Multireq2}.
Specifically, \cite{Mod-Ali-14} and \cite{Mod-Ali-13} study a typical server-user model, where a file server transmits data bits on a shared link (broadcasting channel) to satisfy the request of each user. The objective is to minimize the transmission rate on the shared link. Then, \cite{Mod-Ali-14} and \cite{Mod-Ali-13} propose centralized and decentralized caching
schemes to exploit both local and global caching gains, and thus achieve
multiplicative peak rate reduction on the shared link.
\cite{fundamental4} proposes an online coded caching scheme and achieves the order-optimal long-term performance on the shared link. Besides the single-layer caching in \cite{Mod-Ali-14}-\cite{fundamental4}, \cite{Hier1} proposes an order-optimal file placement and delivery scheme for a two-layer wireless caching network. \cite{Multireq1} generalizes the result in \cite{Hier1} and studies the multi-requests wireless caching networks, where each user requests more than one files. Then, \cite{Multireq1} proposes a caching scheme based on multiple groupcast index coding and achieves the order-optimal performance. \cite{Multireq2} extends the result in \cite{Multireq1} and provides the complete order-optimal characterization of the transmission rate on the shard link.

Considering the popularity diversity of files,
\cite{Nonuniform1}-\cite{Nonuniform4} study nonuniform coded
caching networks, where files are assumed to have different popularity. \cite{Nonuniform1}
considers a wireless caching network with one helper and
multiple users, and develops an order-optimal coded caching scheme. \cite{Nonuniform2}
generalizes the result in \cite{Nonuniform1} to multiple helpers
and users, and optimizes the trade-off among the peak rate on the shared link,
the memory size in helpers, and the access cost of users.
\cite{Nonuniform3} divides the popularity of files into
several discrete levels and derives an information-theoretic outer
bound for the nonuniform network. Different from \cite{Nonuniform3}, where the peak rate on the shared link is considered, \cite{Nonuniform4} optimizes the long-term performance. That is, \cite{Nonuniform4} considers the average rate on the shared link and develops simple order-optimal schemes. Furthermore, \cite{Het_Mem}-\cite{Adaptive} study more practical scenarios and
design effective caching schemes subject to heterogenous cache sizes, delivery delay
constraint, security problem, pricing problem, and streaming schedule problem.

To further improve the system performance, D2D communication is
proposed for wireless caching \cite{D2D1}-\cite{Multihop2}.
\cite{D2D1} considers the D2D communication in wireless caching networks from the
perspective of information theory and proposes a deterministic caching scheme and a random caching scheme, both of which may achieve the information theoretic outer bound within a constant multiplicative factor. Then, \cite{D2D2} and \cite{D2D3} provide the basic principle and system performance of a wireless caching network with D2D communication, and show that the gain from the unicast D2D communication is comparable to the gain from the coded BS multicast. \cite{D2D4} proposes a novel architecture to increase
the system throughput and obtains the optimal collaboration
distance of D2D communication. Different from \cite{D2D1}-\cite{D2D4}, where one-hop D2D communication is allowed, \cite{Multihop1} and \cite{Multihop2} consider multi-hop D2D wireless caching networks, where a requested file can be directly obtained through an one-hop D2D transmission or indirectly obtained after multi-hop D2D transmissions. More specifically, \cite{Multihop1} and \cite{Multihop2} study the throughput scaling law and propose a decentralized caching scheme and a multi-hop D2D transmission scheme. With the schemes, the optimal throughput scaling law is achieved and outperforms the scaling law in one-hop D2D communication networks.

\emph{Contributions:} We adopt the architecture similar to
\cite{D2D4}, where each user may obtain the requested files from its
own memory without any transmission, or from a helper through
an one-hop D2D transmission, or from the BS. However,
our work has three main different aspects from the previous work in \cite{D2D4}. Firstly, we consider a general conventional model, where any two close users are allowed to establish a D2D link, provided that the requested file of one
user is cached in the memory of the other user. This is
different from the cluster-based model in \cite{D2D4}, where a cell
is divided into multiple non-overlapping clusters and only the users
in the same cluster are allowed to establish a D2D link. Thus,
our proposed algorithm has a relaxed constraint
on the location of the users in the cell and can potentially create more D2D communications among users.
Secondly, we manage the mutual interference among different D2D links by efficiently scheduling the D2D links and fairly allocating the transmit power. However, \cite{D2D4} allows at most one D2D link to work in each cluster to avoid strong interference. Thirdly, we consider the max-min fairness of the scheduled D2D links to achieve perfect fairness and study the optimal power allocation at D2D transmitters. This is different from \cite{D2D4}, where the fairness is ignored. As a result, the
three differences lead to a completely different problem formulation
and solution compared with that in \cite{D2D4}. To further clarity, our
contributions are listed as follows.

\begin{itemize}
\item We consider the conventional model, where any two close users are allowed to establish a D2D link provided that the requested file of one user is cached in the memory of the other user. This creates more D2D communications among users compared with the cluster-based model in \cite{D2D4}.

\item We manage the mutual interference among different D2D links by efficiently scheduling the D2D links and fairly allocating the power allocation instead of dividing a cell into non-overlapping clusters and allowing at most one D2D link in each cluster in \cite{D2D4}.

\item We formulate a joint D2D link scheduling and power allocation problem to maximize the system throughput. Due to the non-convexity of the joint optimization problem, we decompose it into a D2D link scheduling problem and an optimal power allocation problem. Specifically, we intend to first maximize the number of the scheduled D2D links by solving the D2D link scheduling problem and then maximize the minimum transmission rate of the scheduled D2D links by solving the optimal power allocation problem.

\item Numerical results indicate that both the number of the scheduled D2D links and the system throughput can be improved simultaneously with the Zipf-distribution caching scheme, the proposed D2D link scheduling algorithm, and the proposed optimal power allocation algorithm compared with the caching schemes and D2D link scheduling algorithms in \cite{D2D3} and \cite{D2D4}.

\end{itemize}

%\newpage
\section{System Model}
\begin{figure}[ht]
            \centering
            \includegraphics[scale=0.5]{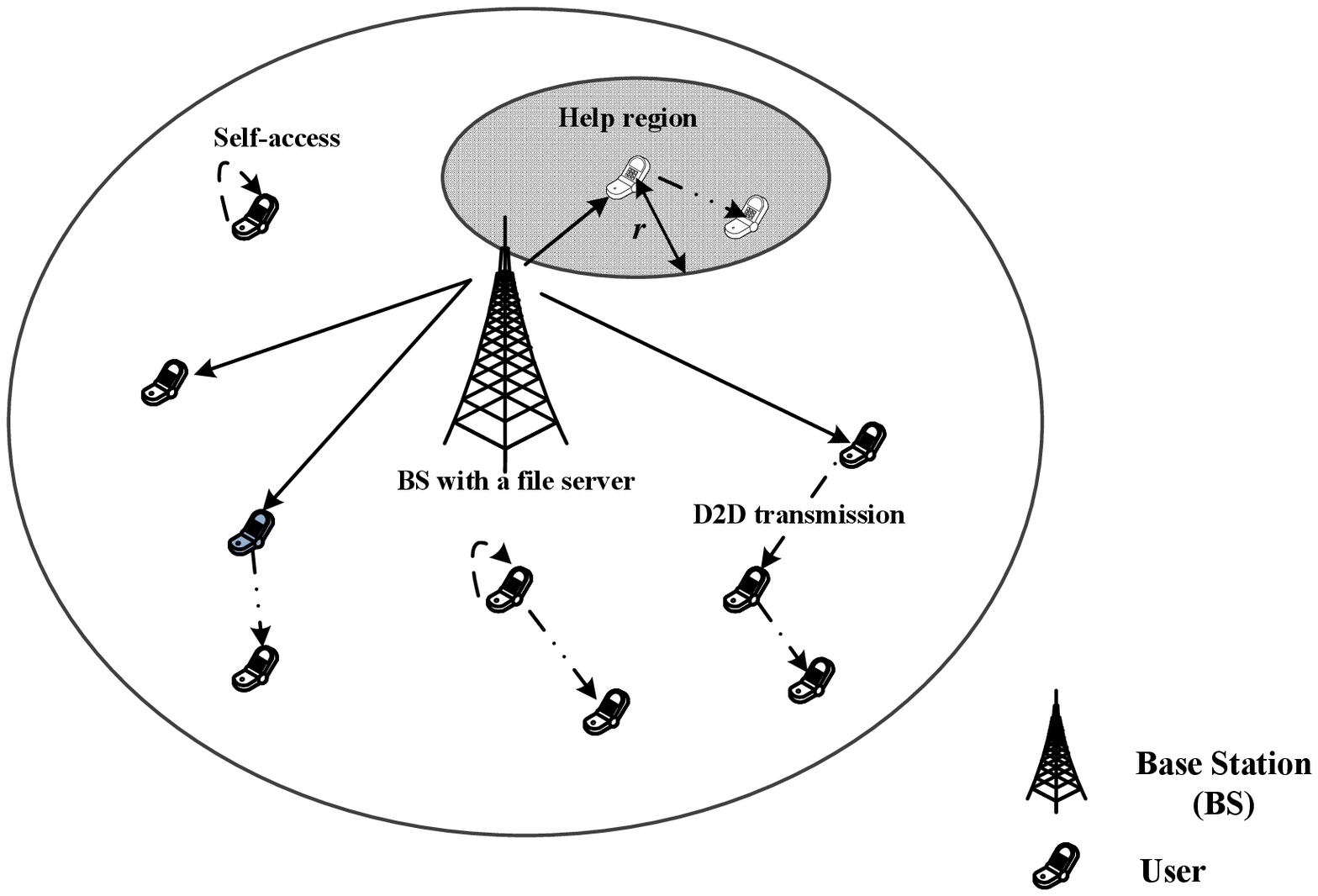}
            \caption{System model, where a user can obtain the requested file from its own memory (dash arrow), or from a helper through an one-hop D2D transmission (dash-dot arrow), or from the BS (solid arrow).}
            \label{system_model}
 \end{figure}

We consider a cellular network with one BS and $K$ users, where the
BS is located at the center of a cell and the users are randomly
distributed in the cell as shown in Fig. \ref{system_model}. The BS has a file sever and stores $N$ files with equal
size\footnote{This can be justified by the fact that big files are
broken into small fragments with the same length.}, whose popularity
probability follows Zipf distribution \cite{ITube}, i.e.,
\begin{equation}
f_{\eta}=\frac{\frac{1}{\eta^{\gamma_r}}}{\sum_{\zeta=1}^{N}\frac{1}{\zeta^{\gamma_r}}},
1 \le \eta \le N, \label{fi}
\end{equation}
where $\eta$ is the file index, and $\gamma_r$ is the file
request coefficient and controls the popularity distribution of
files. Namely, a large $\gamma_r$ means that the first few
files dominate the requests from users.

% \begin{figure}[h]
%            \centering
%            \includegraphics[scale=0.5]{Operation_structure}
%            \caption{Update period structure.}
%            \label{Frame_structure}
% \end{figure}

The file distribution by wireless caching consists of a placement phase and multiple delivery phases.
In the placement phase, each user randomly and independently caches
one out of $N$ files in its memory, according to the Zipf-distribution with a file caching coefficient $\gamma_c$ \footnote{We consider one file to simplify the presentation, although it is straightforward to generalize to multiple files with a Zipf distribution by caching each file independently. Note that the Zipf-distribution caching is not optimal for caching one file or multiple files. We use the Zipf distribution to cache files for four reasons. The first one is that, we focus on the algorithm design in the delivery phase and thus the proposed algorithms in this paper can be used for any caching scheme. The second one is that, the optimal caching distribution of the general model is quite hard to obtain (if it is possible). This is because the optimal caching scheme of one user is related to the number of its neighbors, which is defined as the users in its help region. In \cite{D2D2} and \cite{D2D3}, a cell is divided into multiple non-overlapping clusters and the users in each cluster are assumed to be the neighbors of each other. Then, the numbers of neighbors for all the users in the same cluster are identical. This simplifies the mathematical derivation and makes it possible to obtain the optimal caching distribution for the cluster model. In our general model, the numbers of neighbors in terms of different users are quite different, especially for the users in the center of a cell and the users at the edge of a cell. This makes it quite hard to analyze the optimal caching distribution mathematically for a general model. Thus, a further study on the optimal caching scheme is beyond the scope of this paper. The third one is that, the Zipf distribution can well model the popularity of files \cite{ITube}. Then, we use the Zipf distribution for wireless caching as a heuristic choice. The forth one is that, from the numerical results, the system performance (both the number of the scheduled D2D links and the system throughput) with the Zipf-distribution caching scheme and the proposed algorithms in this paper outperforms the system performance with the optimal caching distribution for the cluster model and the algorithms in \cite{D2D2} and \cite{D2D3}.}. In each delivery phase, each user
requests one file. If the requested file of a user can be found in
its own memory, the user accesses the file without any
transmission. On the other hand, if the user cannot find the
requested file in its own memory, it can request the file
either from a helper through an one-hop D2D transmission or from the BS. Here,
any user in the cell can be a helper of a requesting user if two conditions are satisfied. One is that the two users are \emph{close} in the cell. This is, the requesting user is in the help region,
which is defined by a help distance $r$ around the helper. This may guarantee small
pathloss and small transmit power of a D2D link on average. The other one is that
the requested file is cached in the helper's memory. It should be noted that, a user may
be served by more than one helpers. Then, the helper that can provide the highest
transmission rate or SINR will be chosen as a transmitter in a D2D link. For clarity, the
self-access, the one-hop D2D transmission, and the BS transmission
are illustrated as dash arrow, dash-dot arrow, and solid arrow in
Fig. \ref{system_model}, respectively, and the direction of a arrow
denotes the transmission direction of a file.

We assume that all the D2D links share one WIFI channel and each D2D user works in half-duplex mode and receives or transmits wireless data in one time slot. We also
assume that each user transmits data in the unicast
mode and then cannot transmit data for multiple users
simultaneously. This can be justified by two facts. One is that the asynchronous nature of the requested files, e.g., video clips. This means that the probability that different users request the same video clip from a common helper at the same time is negligible \cite{Nonuniform1}. The other one is that the mobile nature of users. This makes it very hard to adopt the coded caching schemes which exploit the multicast opportunities between one transmitter and multiple receivers in \cite{Mod-Ali-14} and \cite{Mod-Ali-13}. In fact, the request of a video clip in most practical systems is implemented by a point to point (a transmitter to a receiver) transmission with a dedicated connection \cite{D2D2}.

In the following, we will evaluate the SINR at D2D receivers. Suppose that $k_{\text{S}}$ users in set $\mathcal{S}_{\text{S}}$
can access their requested files from their own memory, and
$k_{\text{DB}}$ users in set $\mathcal{S}_{\text{DB}}$ can obtain the
requested files either from helpers through one-hop D2D transmissions or
from the BS, and $k_{\text{B}}$
users in set $\mathcal{S}_{\text{B}}$ can only obtain the requested files
from the BS. Define the link through which a user in $\mathcal{S}_{\text{DB}}$ can obtain the requested file by D2D transmission as a \emph{potential} D2D link and denote $l_m$ as the
potential D2D link with receiver $m$. Without loss of generality, if we
assume $\mathcal{S}_{\text{DB}}=\{1, 2, \cdots, k_{\text{DB}}\}$, the potential D2D
link set can be denoted as $\mathcal{L}_{\text{DB}}=\{l_1, l_2, \cdots,
l_{k_{\text{DB}}}\}$.

Suppose that all the users in $\mathcal{S}_{\text{DB}}$ are scheduled to work simultaneously and that the channel gain between the transmitter of the D2D link $l_m$ ($m \in \mathcal{S}_{\text{DB}}$) to the receiver of the D2D
link $l_n$ ($n \in \mathcal{S}_{\text{DB}}$) is $g(m, n)$. Then, the SINR at the user $m$ (D2D receiver) is
\begin{equation}
v_{m}=\frac{p_{m} g(m, m)}{\sum_{n\in \mathcal{S}_{\text{DB}}, n\neq
m} p_{n} g(n, m)+N_{m}}, \label{gamma_m}
\end{equation}
where $p_{m}$ is the transmit power of the D2D link $l_m$ and it is
subject to the transmit power constraint $0 \leq p_{m} \leq p_m^{\text{max}}$, and $N_{m}$ is the power of the \emph{additive white Gaussian noise} (AWGN) at the receiver $m$. Here, we consider \emph{quality of experience} (QoE) guaranteed D2D transmissions. Thus, the SINR at each D2D receiver is required to be greater than or equal to the minimum acceptable SINR, i.e., $v_m \geq \check{v}_m$, which is related to the modulated and encoding scheme at each D2D transmitter.

%Here, we consider SINR guaranteed D2D link. Thus, the SINR at the receiver of each D2D link is greater than or equals to a
%preset value, i.e., $v_{m} \geq \bar v_{m}$, which is
%related to the file type as well as the modulate and coding scheme
%adopted by $l_m$.
%
%
%
%

\section{Problem Formulation and Analysis}

\subsection{Problem Formulation}
From the previous section, there is a chance that not all the potential D2D links in $\mathcal{L}_{\text{DB}}$ can be scheduled simultaneously for two reasons. One reason is that the potential D2D links in $\mathcal{L}_{\text{DB}}$ may share the same users. That is, one user may be a D2D transmitter in one potential D2D link and a D2D receiver in another potential D2D link. Then, the two links cannot work simultaneously. The other reason is that, some potential D2D links cannot be satisfied with the minimum acceptable SINRs. Thus, we intend to develop a joint D2D link scheduling and power allocation algorithm to maximize the system throughput. Formally, we have
\begin{align}
\nonumber (\text{P}_0): \quad &\underset{p_m: m\in \mathcal{S}_{\text{D}}}{\max} \sum_{m\in \mathcal{S}_{\text{D}}} \log(v_m+1)\\
{{\rm{s}}{\rm{.t}}{\rm{.}}} \ \ & \mathcal{S}_{\text{D}} \subset \mathcal{S}_{\text{DB}},\\
&\mathcal{S}_\text{DT} \cap \mathcal{S}_\text{D}=\varnothing,\\
& v_{m}\geq \check{v}_{m}, \forall \ m\in \mathcal{S}_{\text{D}},\\
  &0 \leq p_m \leq p_m^{\max}, \forall \ m \in \mathcal{S}_{\text{D}},
\label{P0}
\end{align}
where the scheduled D2D link set $\mathcal{S}_{\text{D}}$ and the corresponding transmit power $p_m$, $\forall$ $m\in \mathcal{S}_{\text{D}}$, are the variables to be optimized, $\mathcal{S}_\text{DT}$ is the D2D transmitter set corresponding to the D2D receivers in $\mathcal{S}_\text{D}$, and the constraint $(4)$ guarantees that different scheduled D2D links do not share the same D2D users.

 Unfortunately, the objective function of $(\text{P}_0)$ is non-convex. To obtain the joint optimal D2D link scheduling and power allocation solution, the complexity of the exhaustive search is quite high \footnote{The complexity of the exhaustive search is $O\left(\Pi_{m=1}^{m=k_{\text{DB}}}\frac{p^{\max}_m}{\epsilon}\right)$, where $p^{\max}_m$ is the maximum transmit power of the D2D link $l_m$, $\epsilon$ is the maximum tolerance error of the exhaustive search scheme, and $k_{\text{DB}}$ is the number of users that can obtain the
requested files either from helpers through one-hop D2D transmissions or
from the BS. Suppose that $k_{\text{DB}}=20$, $p^{\max}_m=p^{\max}_n=p^{\max}=100$ mw, $\forall \ m\neq n$, and $\epsilon=0.01$. Then, the complexity of the exhaustive search is extremely large, i.e., $O\left(10000^{20}\right)$.}. Alternatively, we seek to obtain a suboptimal solution of $(\text{P}_0)$ with reasonable complexity. Briefly, we intend to schedule the D2D links with strong communication channels and weak interference channels and allocate the power to the scheduled D2D links fairly. Thus, we decompose $(\text{P}_0)$ into two subproblems\footnote{Note that the decomposition is not optimal and finding the optimal decomposition will be our future work (if it is feasible).}, i.e., a D2D link scheduling problem and an optimal power allocation problem. For the first subproblem, we intend to optimize $\mathcal{S}_{\text{D}}$, to maximize the number of D2D links satisfying both the SINR and the transmit power constraints. For the second subproblem, we intend to optimize $p_m$, $\forall$ $m\in \mathcal{S}_{\text{D}}$, to maximize the minimum transmission rate of the scheduled D2D links. In what follows, we will formulate the two subproblems.

%reduce the traffic load from the BS with wireless caching and maximize the minimum transmission rate of the D. To achieve this, we shall first maximize the number $k_{\text{S}}+k_{\text{D}}$ of users that can obtain the requested files from their own memories or helpers through D2D transmissions. On one hand, $k_{\text{S}}$ is determined by the caching scheme, i.e., the caching coefficient $\gamma_c$. On the other hand, $k_{\text{D}}$ is determined by both the caching scheme and the D2D link scheduling algorithm, which maximizes the number of D2D links satisfying the minimum SINR constraint. Since the affect of the caching scheme and the D2D link scheduling algorithm on the value of $k_{\text{D}}$ are independent, we may develop a D2D link scheduling algorithm irrespective of the caching scheme. In this way, $k_{\text{S}}+k_{\text{D}}$ is only determined by the caching scheme and can be maximized by finding the optimal caching coefficient numerically. After maximizing $k_{\text{S}}+k_{\text{D}}$ with the optimal caching coefficient and the D2D link scheduling algorithm, we obtain the scheduled D2D links. Then, we shall improve the minimum transmission rate of the scheduled D2D links. Thus, we intend to develop an optimal power allocation to maximize the minimum transmission rate of the scheduled D2D links. In what follows, we will first formulate the D2D link scheduling problem and the optimal power allocation problem and then provide the analysis of two problems, respectively.

\subsubsection{D2D Link Scheduling Problem}
To maximize the number of the scheduled D2D links satisfying both the SINR and the transmit power constraints, the D2D link scheduling problem can be written as
\begin{align}
\nonumber (\text{P}_1): \quad &\underset{\mathcal{S}_{\text{D}} \subset \mathcal{S}_{\text{DB}} }{\max}\quad
|\mathcal{S}_{\text{D}}|\\
\nonumber {{\rm{s}}{\rm{.t}}{\rm{.}}}  \ \  & (4), (6),\\
& v_{m}\geq \bar v_m, \forall \ m\in \mathcal{S}_{\text{D}},
\end{align}
where $|\mathcal{S}_{\text{D}}|$ is the cardinality of $\mathcal{S}_{\text{D}}$ and $\bar v_m =\check{v}_{m}$. It is clear that the solution of ($\text{P}_1$) may maximize the number of the scheduled D2D links. However, scheduling the largest number of D2D links may lead to small system throughput since more D2D links may cause more mutual interference. To deal with this issue, we introduce a scheduling coefficient $c_{\text{s}}$ (dB) and replace $\bar v_m =\check{v}_{m}$ with $\bar v_m =\max\{\check{v}_{m}, c_{\text{s}}\}$. Then, if $c_s\leq \underset{m\in \mathcal{S}_{\text{DB}}}{\min} \check{v}_{m}$, we have $\bar v_m =\check{v}_{m}$. Then, ($\text{P}_1$) maximizes the number of the scheduled D2D links. As we increase $c_{\text{s}}$, the SINR constraint (7) becomes strict, only the D2D links with strong communication channels and weak interference channels are scheduled. This potentially increases the system throughput. Thus, we can numerically choose the value of $c_s$ to obtain an acceptable number of the scheduled D2D links as well as the system throughput.

\emph{Theorem 1:} Each transmit power $p_n$, $\forall \ n \in \mathcal{S}_{\text{D}}$, increases if any SINR $v_m$, $ m \in \mathcal{S}_{\text{D}}$, in $(\text{P}_1)$ increases. Meanwhile, the value of $|\mathcal{S}_{\text{D}}|$ remains constant or decreases if any SINR $v_m$, $ m \in \mathcal{S}_{\text{D}}$, in $(\text{P}_1)$ increases.

\begin{proof}
The proof is provided in Appendix A.
\end{proof}

From Theorem 1, the value of $|\mathcal{S}_{\text{D}}|$ may decrease if any SINR $v_m$ in $\mathcal{S}_{\text{D}}$ increases. Then, problem $(\text{P}_1)$ is equivalent to
\begin{align}
\nonumber (\text{P}_2): \quad &\underset{\mathcal{S}_{\text{D}} \subset \mathcal{S}_{\text{DB}} }{\max}\quad
|\mathcal{S}_{\text{D}}|\\
\nonumber & {{\rm{s}}{\rm{.t}}{\rm{.}}}  \ \ (4), (6),\\
 &\quad\quad v_{m} = \bar v_{m}, \forall \ m \in \mathcal{S}_{\text{D}}.
\end{align}

By solving problem $(\text{P}_2)$, we may obtain the optimal D2D receiver set $\mathcal{S}^*_{\text{D}}$ and the corresponding D2D link set $\mathcal{L}^*_{\text{D}}$. Without loss of generality, we assume $\mathcal{S}^*_{\text{D}}=\{1, 2, \cdots,
k_{\text{D}}\}$ and $\mathcal{L}^*_{\text{D}}=\{l_1, l_2, \cdots,
l_{k_{\text{D}}}\}$. It is clear that problem $(\text{P}_2)$ outputs the SINR vector $\mathbf{\bar V}=[\bar v_1, \bar v_2, \cdots, \bar v_{k_{\text{D}}}]^{\text{T}}$ at the scheduled D2D receivers. Accordingly, we assume the power allocation vector is $\mathbf{\bar P}=[\bar p_1, \bar p_2, \cdots, \bar
p_{k_{\text{D}}}]^{\text{T}}$.

\subsubsection{Optimal Power Allocation Problem} From Theorem 1 again, the value of $|\mathcal{S}^*_{\text{D}}|$ may remain constant
if any SINR $v_m$, $m\in\mathcal{S}^*_{\text{D}}$, increases. Thus,
we may improve the minimum transmission rate or the minimum SINR of the scheduled D2D links
without compromising $|\mathcal{S}^*_{\text{D}}|$, provided that the transmit power constraints
of the scheduled D2D links are not violated, i.e.,
\begin{align}
\nonumber (\text{P}_3): \quad &\underset{p_m:m \in \mathcal{S}^*_{\text{D}}}{\max} {\underset{1 \leq m \leq k_{\text{D}}}{\min} \log(v_m+1)}\\
{{\rm{s}}{\rm{.t}}{\rm{.}}} \ \ & v_{m}\geq \bar v_{m}, \forall \ m\in \mathcal{S}^*_{\text{D}},\\
  &\bar p_m \leq p_m \leq p_m^{\max}, \forall \ m \in \mathcal{S}^*_{\text{D}}.
\label{P3}
\end{align}

\subsection{Analysis of Subproblem $(\text{P}_2)$}
To solve problem $(\text{P}_2)$, we shall develop a D2D link scheduling algorithm
to select the largest number of D2D links satisfying all the constraints in problem $(\text{P}_2)$. In fact, problem $(\text{P}_2)$ is an admission
control problem (i.e., link scheduling problem) \cite{DCPC}-\cite{Fair_Throughput}. However, it is different from a regular admission control problem. In the regular admission control problem, one transmitter has one dedicated receiver. In our system model,
there is a chance that one user is the transmitter in one potential D2D link and the receiver of another potential D2D link. Since the scheduled D2D links cannot share the same D2D users, i.e.,
the constraint $(4)$, problem $(\text{P}_2)$ is more complicate than the
regular admission control problem and the algorithm for the regular admission
control problem cannot be directly used to solve problem $(\text{P}_2)$. Thus, we
need to develop a scheduling algorithm to solve problem $(\text{P}_2)$.

Denote the events that any two potential D2D links in a potential D2D link set do not
share the same users and that all the minimum SINR constraints in
a potential D2D link set can be satisfied
as $C_1$ and $C_2$. Accordingly, $\bar C_1$ and
$\bar C_2$ denote the events that $C_1$ and $C_2$ cannot be
satisfied, respectively. Then, we have the following three cases
depending on whether $C_1$ and/or $C_2$ can be satisfied in the potential D2D link set
$\mathcal{L}_{\text{DB}}$.

Case I: $(C_1, C_2)$. In this case, all the potential D2D links in
$\mathcal{L}_{\text{DB}}$ will be scheduled. Then, we have $\mathcal{S}^*_{\text{D}}=\mathcal{S}_{\text{DB}}$.

Case II: $(C_1, \bar C_2)$. In this case, problem $(\text{P}_2)$ reduces to a
regular admission control problem. To solve this problem, the fewest
D2D links in $\mathcal{L}_{\text{DB}}$ shall be removed until $C_2$ can be
satisfied.

Case III: $(\bar C_1)$. In this case, some D2D links share the same users and are not allowed to work simultaneously. To solve this problem, we
shall first divide the potential D2D links in $\mathcal{L}_{\text{DB}}$ into different
groups (or subsets), such that any two potential D2D links in one group do not
share the same users. After that, each group becomes case I or case
II.

For generality, we will consider Case III for
problem $(\text{P}_2)$ in the rest of this paper. By analyzing problem $(\text{P}_2)$, we have

\emph{Theorem 2:} Problem $(\text{P}_2)$ is
NP-hard.

\begin{proof}
The proof is provided in Appendix B.
\end{proof}

Considering the highly computational complexity of obtaining the
optimal solution for a NP-hard problem, we will develop an efficient
algorithm to obtain a suboptimal solution of problem
$(\text{P}_2)$ in what follows.

\subsection{Analysis of Subproblem $(\text{P}_3)$}

Once the optimal D2D set $\mathcal{S}^*_{\text{D}}$ is obtained from problem
$(\text{P}_2)$, problem $(\text{P}_3)$ is a feasible max-min
optimization problem with transmit power of D2D links as
variables. Then, we will first analyze the property of the optimal
power allocation of problem $(\text{P}_3)$ and then develop an optimal
power allocation algorithm to achieve the optimal performance.

\section{D2D link Scheduling}
In this section, we aim to maximize the number of the
scheduled D2D links. However, some of the potential D2D
links may be dependent. That is, different potential
D2D links in $\mathcal{L}_{\text{DB}}$ share the same users. This is
the main difference between problem $(\text{P}_2)$
and a regular admission control problem, where any two different potential D2D
links do not share the same users. Thus, we shall divide
the potential D2D links in $\mathcal{L}_{\text{DB}}$ into different
groups, such that different potential D2D links in each group are
independent. In this way, problem $(\text{P}_2)$
is decomposed into several subproblems corresponding to different
groups. After maximizing the number of the potential D2D links
that satisfying both $C_1$ and $C_2$ for each subproblem (group), we
can obtain a suboptimal solution of problem $(\text{P}_2)$.

In what follows, we will first divide the potential D2D
links in $\mathcal{L}_{\text{DB}}$ into different groups such that the
potential D2D links in each group are independent and
satisfy $C_1$. Then, we will maximize the number of the potential
D2D links satisfying $C_2$ in each group. Specifically, we develop a
\emph{centralized power control} (CPC) algorithm to check the state
of each group, i.e., to decide whether the potential D2D links in each
group satisfy $C_2$ or not. If the potential D2D links in one
group do not satisfy $C_2$, we further develop a removal algorithm
to remove the fewest potential D2D links from this group.
After applying the CPC algorithm and/or the removal algorithm into
each group, the largest number of the potential D2D links
satisfying both $C_1$ and $C_2$ in each group is obtained.
Finally, the potential D2D links in the group with the
largest number of potential D2D links will be scheduled and
other potential D2D links will not be allowed.

\subsection{Potential D2D Links Division}

To maximize the number of the scheduled D2D links, we shall
add as many potential D2D links as possible into each group. This leads
to the fewest groups. Thus, the division of the potential D2D links can be
transferred to an edge coloring problem in graph theory, where the
edges in a graph are colored with the fewest colors while guaranteeing
that any two edges sharing the same vertex are colored
differently.

Denote the connection among the users involved the potential D2D links in
$\mathcal{L}_{\text{DB}}$ as a graph $G(\mathcal{O}, \mathcal{E})$, where $\mathcal{O}$ is the vertex set
denoting the involved users in $\mathcal{L}_{\text{DB}}$, $\mathcal{E}$ is the edge set
denoting the D2D links in $\mathcal{L}_{\text{DB}}$. Then, we may color the
edges in $\mathcal{E}$ with $\chi^{'}(G)$ colors, which is the edge
chromatic number \cite{GT}. Since the edge coloring problem is NP-hard for a general graph, it is non-trivial to obtain the optimal solution. Alternatively, we develop an iterative edge-coloring algorithm as shown in \textbf{Algorithm \ref{Iterative_coloring}}, which colors the edges in $\mathcal{E}$ with $N_{\text{C}}$ colors ($N_{\text{C}}\geq \chi^{'}(G)$). More formally, denote
$\mathcal{E}=\mathcal{E}_1 \cup ... \cup \mathcal{E}_{N_{\text{C}}}$, where $\mathcal{E}_i$ ($1 \leq i \leq
N_{\text{C}}$) represents the edge set with the $i$th color after edge
coloring. We assume that there are $k_i$ edges in $\mathcal{E}_i$, which
corresponds to $k_i$ potential D2D links. If we denote the receiver set of the
$k_i$ potential D2D links as $\mathcal{S}_i=\{i_1, i_2, \cdots, i_{k_i}\}$, the $k_i$ potential D2D link set can be denoted as $\mathcal{L}_i=\{l_{i_1}, l_{i_2}, \cdots,
l_{i_{k_i}}\}$.

In the $i$th iteration of \textbf{Algorithm \ref{Iterative_coloring}}, the maximum matching \footnote{The maximum matching is the maximum edge set, where any two edges do not share a common vertex.} is obtained by solving a linear programming problem \cite{LP}
\begin{align}
\nonumber (\text{P}_4): \quad &\underset{x_e: e\in \mathcal{E}}{\max} \ \sum_{e\in \mathcal{E} } \ x_e\\
{{\rm{s}}{\rm{.t}}{\rm{.}}} \quad  &\sum_{e \sim o } \ x_e \leq 1, \forall \ o \in \mathcal{O},\\
& 0\leq x_e \leq 1, \forall \ e\in \mathcal{E} ,
\label{P4}
\end{align}
where $e \sim o$ means that the edge $e$ is incident on the vertex $o$, and rounding the solutions $x_e$ $(e \in \mathcal{E})$ to $0$ or $1$ in order to map the solutions to integers. That is, $x_e=0$ if $0 \leq x_e < 0.5$ and $x_e=1$ if $0.5 \leq x_e \leq 1$. In this way, the edges $x_e=1$ are colored with the $i$th color and put into the set $\mathcal{L}_i=\{l(e): x_e=1, e \in \mathcal{E}\}$, where $l(e)$ denotes the D2D link corresponding to the edge $e$. This procedure is terminated until all the edges are colored. Note that the computational complexity of \textbf{Algorithm \ref{Iterative_coloring}} is dominated by solving the linear programming problem $(\text{P}_4)$ with computational complexity $O(k_{\text{DB}}^3)$ \cite{Convex}. Thus, the overall computational complexity of \textbf{Algorithm \ref{Iterative_coloring}} is $O(k_{\text{DB}}^3)$.

\begin{algorithm}[htb]

\algsetup{linenosize=\footnotesize }

\caption{Iterative Edge-coloring Algorithm.}

\label{Iterative_coloring}

\begin{algorithmic}[1]

\REQUIRE ~~\ \

\STATE $G(\mathcal{O}, \mathcal{E})$; i=0;

\ENSURE ~~\ \

\WHILE{$\mathcal{E}\neq \emptyset$}

\STATE $i=i+1$, $\mathcal{L}_i=\emptyset$;

\STATE Solve $(\text{p}_4)$ and obtain $x_e$, $\forall \ e \in \mathcal{E}$, and round $x_e$ to $0$ or $1$;

\STATE Color all the edges $x_e=1$ with the $i$th color;

\STATE $\mathcal{L}_i=\{l(e): x_e=1, e \in \mathcal{E}\}$, $\mathcal{E}= \mathcal{E} \backslash \{e: x_e=1\}$, ;

\ENDWHILE
\end{algorithmic}

\end{algorithm}

\subsection{CPC Algorithm}
After dividing the potential D2D links into $\mathcal{L}_i$ ($1 \leq i \leq
N_{\text{C}}$) with the iterative edge-coloring algorithm, $C_1$ is satisfied among the potential D2D links in each $\mathcal{L}_i$. In what follows, we will develop an algorithm to check the state of
the potential D2D links in $\mathcal{L}_i$, i.e., to decide whether the potential D2D links in $\mathcal{L}_i$ satisfy
$C_2$ or not.

If all the potential D2D links in $\mathcal{L}_i$ satisfy $C_2$, there exists a power allocation to satisfy all the potential D2D links in $\mathcal{L}_i$ with the minimum SINR constraints $ \mathbf{\bar V}_{i}=[\bar v_{i_1}, \ \bar
v_{i_2},\ ..., \ \bar v_{i_{k_i}}]^{\text{T}}$. Suppose that the power allocation is $\mathbf{\bar P}_{i}=[\bar p_{i_1}, \ \bar p_{i_2},\ ...,
\ \bar p_{i_{k_i}}]^{\text{T}}$. Then, we have
\begin{equation}
\emph{H}_i(\mathbf{\bar V}_{i})\mathbf{\bar P}_i=\mathbf{N}_i,
\end{equation}
where
\begin{equation}
\emph{H}_i(\mathbf{\bar V}_{i}) \!\!=\!\! \left[ {\begin{array}{*{20}c}
   \!\!{g(i_1, i_1)/ \bar v_{i_1} } & \!\!\!\!\! { -g(i_1, i_2) } & \!\!\!\!\!  \cdots \!\!\!\!\! & {  -g(i_1,i_{k_i}) } \!\!\!\!\! \\
   \!\!\!\!\!{ -g(i_2, i_1) } & \!\!\!\!\! {g(i_2, i_2)/ \bar v_{i_2} } & \!\!\!\!\! \cdots \!\!\!\!\! & { -g(i_2, i_{k_i}) } \!\!\!\!\! \\
    \!\!\!\!\! \vdots  & \!\!\!\!\! \vdots  &  \ddots  & \!\!\!\!\! \vdots \!\!\!\!\! \!\!\!\!\! \\
   \!\!\!\!\!{ -g(i_{k_i},i_1) } & \!\!\!\!\! { -g(i_{k_i}, i_2) } & \!\!\!\!\! \cdots \!\!\!\!\! & {g(i_{k_i}, i_{k_i})/\bar v_{i_{k_i}} } \!\! \\
\end{array}} \right]^{\text{T}}
\label{Define_H}
\end{equation}
and
\begin{equation}
\mathbf{N}_i = \left[N_{i_1}, \ N_{i_2}, \ \cdots, \ N_{i_{k_i}}
\right]^{\text{T}}.
\end{equation}

%\emph{Theorem 2:} The power
%allocation $p_{i_m}$ is non-decreasing in terms of the
%minimum SINR requirement $\gamma_{i_m}$ $\forall$ $i_m \in \Omega_i$.
%
%\begin{proof}
%The Theorem can be easily obtained from (\ref{optimal_equation}).
%\end{proof}

Then, the power allocation $\mathbf{\bar P}_i$ can be derived as
\begin{equation}
\mathbf{\bar P}_i=\emph{H}_i^{-1}(\mathbf{\bar V}_{i})\mathbf{N}_i.
\label{P_star}
\end{equation}

If the potential D2D links in $\mathcal{L}_i$ satisfy $C_2$, we have $\mathbf{0} \preceq
\mathbf{\bar P}_i \preceq
\mathbf{P}_i^{\text{max}}$, where $\mathbf{P}_i^{\max}=[p_{i_1}^{\max}, p_{i_2}^{\max},
\cdots, p_{i_{k_i}}^{\max}]^{\text{T}}$, $\mathbf{0}$ is a $k_i
\times 1$ vector with all zero elements, and ``$\preceq$'' is an element-wise operator.

Consequently, we have the CPC algorithm as
\begin{itemize}
\item the potential D2D links in $\mathcal{L}_i$ satisfy $C_2$ if $\mathbf{0} \preceq
\mathbf{\bar P}_i \preceq
\mathbf{P}_i^{\text{max}}$ holds.
\item the potential D2D links in $\mathcal{L}_i$ do not satisfy $C_2$ if $\mathbf{0} \preceq
\mathbf{\bar P}_i \preceq
\mathbf{P}_i^{\text{max}}$ does not
holds.
 \end{itemize}

In our algorithm, the BS first collects the channel gains
among different potential D2D links and the minimum SINR constraint of each
potential D2D link, and then calculates the required power allocation to
satisfy the potential D2D links in $\mathcal{L}_i$ with their minimum SINR constraints.
Thus, we name this algorithm as centralized power control (CPC) algorithm. Essentially, the main ideas of the CPC algorithm and the \emph{distributed constrain power control} (DCPC) algorithm in \cite{DCPC} are similar: check whether multiple transceivers can be supported simultaneously by
checking whether the transmit power constraints are violated to
achieve the minimum SINR constraints. However, the CPC algorithm is
more efficient than the DCPC algorithm to implement. More specifically, to
implement the DCPC algorithm, each D2D transmitter in $\mathcal{L}_i$ iteratively
adapts its transmit power to achieve the minimum SINR
constraint until all the transmit power converges. If all the minimum SINR constraints are satisfied after
transmit power converges, the potential D2D links in $\mathcal{L}_i$ satisfy $C_2$. If
some minimum SINR constraints are not satisfied, the potential D2D links in $\mathcal{L}_i$ do not satisfy $C_2$. Then, a removal algorithm will be developed to remove some
potential D2D links from $\mathcal{L}_i$. In the CPC algorithm, the BS first collects
all the necessary information and then calculates the power
allocation to achieve the minimum SINR constraints. If the
calculated power allocation can be satisfied at all the D2D transmitters
in $\mathcal{L}_i$, the potential D2D links in $\mathcal{L}_i$ satisfy $C_2$. Otherwise, a removal
algorithm will be developed.

\subsection{Removal Algorithm}
From the CPC algorithm, there exists at least one calculated transmit power $\bar p_{i_m}$ in $\mathcal{L}_i$
satisfying
$\bar p_{i_m}>p_{i_m}^{\text{max}}$ or $\bar p_{i_m}<0$ if the potential D2D
links in $\mathcal{L}_i$ do not satisfy $C_2$. This is because the mutual
interference among the potential D2D links in $\mathcal{L}_i$ is too strong. Then, we
shall remove some potential D2D links from $\mathcal{L}_i$ to enable the remaining ones
to satisfy $C_2$. Our approach is to remove the potential D2D link which is likely
to cause the strongest interference to others or receive the strongest interference from others
each time until the remaining potential D2D links satisfy $C_2$.

Specifically, to satisfy the minimum SINR constraint $\bar v_{i_m}$ ($i_m
\in \mathcal{S}_i$), the transmitter of the potential D2D link $l_{i_m}$ has to set
the transmit power no less than $\frac{N_{i_m} \bar
v_{i_m}}{g(i_m,i_m)}$ and causes no less than $\frac{N_{i_m} \bar
v_{i_m}}{g(i_m,i_m)}g(i_m,i_n)$ ($i_n \in \mathcal{S}_i, i_n \neq
i_m$) interference to another potential D2D link $l_{i_n}$. Since a potential D2D link with a smaller minimum SINR constraint and a larger maximum transmit power can
tolerate more interference from other potential D2D links, we define the \emph{relative
interference} from $l_{i_m}$ to $l_{i_n}$ as
$I_r(l_{i_m},l_{i_n})=\frac{\bar
v_{i_n}}{p_{i_n}^{\max}}\frac{N_{i_m} \bar
v_{i_m}}{g(i_m,i_m)}g(i_m,i_n)$. Then, the summation of relative
interference generated by $l_{i_m}$ is larger than
\begin{equation}
\alpha_{i_m}=\sum_{n=1, n\neq
m}^{n=k_i} I_r(l_{i_m},l_{i_n})=\frac{N_{i_m} \bar
v_{i_m}}{g(i_m,i_m)}\sum_{n=1, n \neq
m}^{n=k_i}\frac{\bar v_{i_n}}{p_{i_n}^{\max}}g(i_m,
i_n).
\end{equation}
Similarly, to satisfy the minimum SINR constraint $\bar v_{i_n}$ ($i_n \in
\mathcal{S}_i, i_n \neq i_m$), the transmitter of the potential D2D link $l_{i_n}$
has to set the transmit power no less than $\frac{N_{i_n} \bar
v_{i_n}}{g(i_n,i_n)}$. Then, the summation of relative
interference to the potential D2D link $l_{i_m}$ is larger than
\begin{equation}
\beta_{i_m}=\sum_{n=1, n \neq m}^{n=k_i} I_r(l_{i_n},l_{i_m})=
\frac{\bar v_{i_m}}{p_{i_m}^{\max}} \sum_{n=1, n \neq
m}^{n=k_i}\frac{N_{i_n} \bar v_{i_n}}{g(i_n,i_n)}g(i_n, i_m).
\end{equation}

Thus, the potential D2D link $l_{i_{m^*}}$, where
\begin{equation}
i_{m^*}=\underset{1\leq m \leq k_i}{\text{arg max}}\ \
 \max\{\alpha_{i_m}, \beta_{i_m}\} \label{removal_2}
\end{equation}
is likely to cause the strongest interference to others or receive the strongest interference from others and will be removed from $\mathcal{L}_i$.

It should be noted that, the proposed removal algorithm is different
from the DCPC-based removal algorithm in \cite{DCPC}. Specifically,
our proposed algorithm utilizes the information of the maximum
transmit power at each transmitter while the DCPC-based removal
algorithm in \cite{DCPC} utilizes the converged transmit power of the
DCPC algorithm, although the two algorithms share other two kinds
of information, i.e., the minimum SINR constraints and the interference channel gains among different potential D2D links. Besides, our proposed removal
algorithm depends on the relative interference among different transmissions
whereas the DCPC-based removal algorithm in \cite{DCPC} is developed
by measuring the absolute interference after the DCPC algorithm converges.
Thus, the calculated interference in our proposed algorithm is more accurate than that in \cite{DCPC} and our proposed removal algorithm is more flexible than the
DCPC-based removal algorithm in \cite{DCPC}. In fact, from the numerical results in Section VI, our proposed removal algorithm outperforms the
DCPC-based removal algorithm.

\subsection{Scheduling Algorithm and Complexity Analysis}
After applying the CPC algorithm and/or removal algorithm into each potential D2D link set
$\mathcal{L}_i$ ($1 \leq i \leq
N_{\text{C}}$), the potential D2D links in each $\mathcal{L}_i$ satisfy both $C_1$ and $C_2$.
Then, the potential D2D links in $\mathcal{L}_{i^*}=\underset{1 \leq i
\leq N_{\text{C}}}{\text{arg max}} \ |\mathcal{L}_i|$ with the largest number of the potential D2D links
will be scheduled. To summarize, we illustrate the detail D2D link scheduling
algorithm in \textbf{Algorithm \ref{Schedule}}. 

Note that \textbf{Algorithm \ref{Schedule}} consists of \textbf{Algorithm \ref{Iterative_coloring}}, the CPC algorithm, and the removal algorithm. The computational complexity of \textbf{Algorithm \ref{Iterative_coloring}} is $O(k_{\text{DB}}^3)$. The computational complexity of the CPC algorithm is dominated by calculating (\ref{P_star}), i.e., $\mathbf{\bar P}_i=\emph{H}_i^{-1}(\mathbf{\bar V}_{i})\mathbf{N}_i$. In the worst case, $\emph{H}_i(\mathbf{\bar V}_{i})$ is a $k_{\text{DB}}\times k_{\text{DB}}$ matrix and $\mathbf{N}_i$ is a $k_{\text{DB}}\times 1$ vector. The computational complexity of the inverse of a $k_{\text{DB}}\times k_{\text{DB}}$ matrix $\emph{H}_i(\mathbf{\bar V}_{i})$ is $O(k_{\text{DB}}^{2.373})$ and the computational complexity of the multiplication of a $k_{\text{DB}}\times k_{\text{DB}}$ matrix $\emph{H}_i^{-1}(\mathbf{\bar V}_{i})$ and a $k_{\text{DB}}\times 1$ vector $\mathbf{N}_i$ is $O(k_{\text{DB}}^2)$ \cite{Complexity_2}. Then, the computational complexity of the CPC algorithm is $O(k_{\text{DB}}^{2.373})+O(k_{\text{DB}}^2)=O(k_{\text{DB}}^{2.373})$. Besides, the computational complexity of the removal algorithm is dominated by arithmetic and thus is $O(k_{\text{DB}})$. Consequently, the computational complexity of \textbf{Algorithm \ref{Schedule}} is $O(k_{\text{DB}}^3)+O(k_{\text{DB}}^{2.373})+O(k_{\text{DB}})=O(k_{\text{DB}}^3)$.

\begin{algorithm}[htb]

\algsetup{linenosize=\footnotesize }

\caption{D2D Link Scheduling Algorithm.}

\label{Schedule}

\begin{algorithmic}[1]

\REQUIRE ~~\ \

\STATE $G(\mathcal{O}, \mathcal{E})$;

\ENSURE ~~\ \

\STATE Execute Algorithm \ref{Iterative_coloring} to color $\mathcal{E}$ with $N_{\text{C}}$ colors,
i.e., $\mathcal{E}=\mathcal{E}_1 \cup ... \cup \mathcal{E}_{N_{\text{C}}}$ ;

\FOR{$i=1$ to $i=N_{\text{C}}$}

\STATE Adopt CPC algorithm to check whether the potential D2D links in $\mathcal{L}_i$ corresponding to the edge set $\mathcal{E}_i$ satisfy $C_2$;

\IF{the potential D2D links in $\mathcal{L}_i$ do not satisfy $C_2$} \WHILE{the potential D2D links in
$\mathcal{L}_i$ do not satisfy $C_2$}

\STATE Remove the potential D2D link $i_{m^*}$ according to (\ref{removal_2});

\STATE Update $\mathcal{L}_i= \mathcal{L}_i \backslash l_{i_{m^*}}$;

\ENDWHILE \ENDIF \ENDFOR

\STATE BS schedules the potential D2D links in set $\mathcal{L}_{i^*}=\underset{1 \leq i \leq
N_{\text{C}}}{\text{arg max}} \ |\mathcal{L}_i|$.
\end{algorithmic}

\end{algorithm}
%
%\subsection{Implementation Issues}
%The proposed schedule algorithm is performed by the BS in a
%centralized way and requires the channel gain information between
%each D2D transmitter and each D2D receiver, i.e., $h_i(m,n)$,
%$\forall \ i, \ m, \ n$, the minimum SINR constraints $\bar
%\gamma_{i,m}$ of each D2D link, and the AWGN power $N_{i, m}$ at
%each D2D receiver. Actually, these information can be collected by
%the BS. Briefly speaking, the the minimum SINR constraint can be
%obtained by the BS after identifying the specific service. The AWGN
%power can be estimated by the D2D receiver and forwarded to the BS.
%To obtain the channel information, D2D transmitters send probing
%signal for channel estimation and D2D receivers estimate the
%channels and forward them to the BS \cite{Channel_1}\cite{Channel_2}.

\section{Optimization of Power Allocation}
In the previous section, we have solved problem $(\text{P}_2)$ and obtained the scheduled D2D link set $\mathcal{L}_{i^*}=\mathcal{L}^*_{\text{D}}=\{l_1, l_2, \cdots,
l_{k_{\text{D}}}\}$ and the D2D receiver set $\mathcal{S}^*_{\text{D}}=\{1, 2,
\cdots, k_{\text{D}}\}$. Accordingly, we may obtain the output transmit power vector
and the SINR vector as $\mathbf{\bar P}_{i^*}=\mathbf{\bar P}=[\bar p_1, \bar p_2, \cdots,
\bar p_{k_{\text{D}}}]^{\text{T}}$ and $\mathbf{V}_{i^*} =\mathbf{\bar V}=[\bar v_1, \bar v_2, \cdots,
\bar v_{k_{\text{D}}}]^{\text{T}}$ from problem $(\text{P}_2)$. In this section, we will
solve problem $(\text{P}_3)$ to maximize the minimum transmission rate or the minimum SINR of the scheduled D2D links without
compromising the number of the scheduled D2D links. Although
problem $(\text{P}_3)$ can be transformed to a convex
problem and is solved with a convex optimization toolbox, we seek to
solve it analytically to shed more light on the optimal power
allocation. Specifically, we will first analyze the property of the
optimal solution of problem $(\text{P}_3)$. Then, we will develop a
binary-search based power allocation to
obtain the optimal power allocation. Finally, we analyze the
computational complexity of our algorithm.

\subsection{Binary-search based Power Allocation}

 As mentioned above, the minimum SINR constraints $\mathbf{V}=\mathbf{\bar V}=[\bar v_1,
\bar v_2, \cdots, \bar v_{k_{\text{D}}}]^{\text{T}}$ of
the scheduled D2D links are satisfied with the power allocation
$\mathbf{\bar P}=[\bar p_1, \bar p_2, \cdots, \bar
p_{k_{\text{D}}}]^{\text{T}}$. To further increase the SINRs in $\mathbf{V}$ and maximize
the minimum SINR in $\mathbf{V}$, i.e., problem $(\text{P}_3)$, we shall increase
the SINRs in $\mathbf{V}$ with small minimum SINR constraints with priority.

Intuitively, if we assume $\bar v_1 \leq \bar v_2 \leq
\cdots \leq \bar v_{k_{\text{D}}}$, we shall increase
$v_1$ from $\bar v_1$ to $\bar v_2$, i.e., $\bar
v_1 \leq v_1 \leq \bar v_2$, and then increase
$v_1$ and $v_2$ from $\bar v_2$ to $\bar v_3$,
i.e., $\bar v_2 \leq v_1=v_2 \leq \bar v_3$, and
then increase $v_1$, $v_2$, $\cdots$, $v_m$ ($1 \leq
m<{k_{\text{D}}}$) from $\bar v_m$ to $\bar v_{m+1}$,
i.e., $\bar v_m \leq v_1=v_2=\cdots=v_m \leq
\bar v_{m+1}$. This procedure is terminated until some maximum
transmit power constraints are violated. In this way, we
maximize the minimum SINR of the scheduled D2D links. More
formally, we have the following Theorem.

\emph{Theorem 3:} Assume $\bar v_1 \leq \bar
v_2 \leq \cdots \leq \bar v_{k_{\text{D}}}$ and denote $\mathbf{V}^{(m)}(
v)=[\underbrace {
v, \cdots,  v}_{m}, \bar v_{m+1}, \cdots,
\bar v_{k_{\text{D}}}]$, we have $\mathbf{V}^{(m)}(\bar
v_m)=[\underbrace {\bar
v_m, \cdots, \bar v_m}_{m}, \bar v_{m+1}, \cdots,
\bar v_{k_{\text{D}}}]$ and $\mathbf{P}(\mathbf{V}^{(m)}(\bar
v_m))=\emph{H}^{-1}(\mathbf{V}^{(m)}(\bar
v_m))\mathbf{N}$, where $\emph{H}(\mathbf{V}^{(m)}(\bar
v_m))$ is given in (\ref{long_equation_1}) and $\mathbf{N} =
\left[N_{1}, \ N_{2}, \ \cdots, \ N_{k_{\text{D}}}
\right]^{\text{T}}$.

\begin{figure*}[t!]
\begin{eqnarray}
\begin{split}
&\emph{H}(\mathbf{V}^{(m)}(\bar
v_m))=\\
&\left[ {\begin{array}{*{20}c}
   {\!\!\!g(1,1)/\bar v _m } \!\!\!& { - g(1,2)} \!\!\!&  \!\!\!\cdots  \!\!\!& { - g(1,m)} \!\!\!&\!\!\! { - g(1,m \!+\! 1)} \!\!\!&\!\!\!  \cdots  \!\!\!&\!\!\! { - g(1,k_{{\rm{D}}} )}  \\
   { \!\!\!- g(2,1)} \!\!\!& {g(2,2)/\bar v _m } \!\!\!&  \!\!\!\cdots  \!\!\!& { - g(2,m)} \!\!\!&\!\!\! { - g(2,m \!+\! 1)} \!\!\!&\!\!\!  \cdots  \!\!\!&\!\!\! { - g(2,k_{{\rm{D}}} )}  \\
    \!\!\!\vdots  \!\!\!&\!\!\!  \vdots  \!\!\!&\!\!\!  \ddots  \!\!\!&  \!\!\!\vdots  \!\!\!&  \!\!\!\vdots  \!\!\!&\!\!\!  \ddots  \!\!\!&\!\!\!  \vdots   \\
   {\!\!\! - g(m,1)} \!\!\!&\!\!\! { - g(m,2)} \!\!\!&  \!\!\!\cdots  \!\!\!&\!\!\! {g(m,m)/\bar v _m } \!\!\!&\!\!\! { - g(m,m \!+\! 1)} \!\!\!&  \!\!\!\cdots  \!\!\!& \!\!\!{ - g(m,k_{{\rm{D}}} )}  \\
   { \!\!\!- g(m \!\!+\!\! 1,1)} \!\!&\!\! { - g(m \!\!+\!\! 1,2)}  \!\!& \!\!\!\cdots  \!\!\!&\!\! { - g(m \!+\! 1,m)}  \!&\! {g(m \!+\! 1,m \!+\! 1)/\bar v _{m \!+\! 1} } \!\!\!&\!\!\!  \cdots \!\!\! &\!\!\! { - g(m \!+\! 1,k_{{\rm{D}}} )}  \\
    \!\!\!\vdots  &  \!\!\!\vdots  &  \!\!\!\ddots  &  \!\!\!\vdots  &  \vdots  &  \ddots  &  \vdots   \\
   { \!\!\!- g(k_{{\rm{D}}} ,1)} \!\!\!&\!\!\! { - g(k_{{\rm{D}}} ,2)} \!\!\!&  \!\!\!\cdots  \!\!\!&\!\!\! { - g(k_{{\rm{D}}} ,m)} \!\!\!&\!\!\! { - g(k_{{\rm{D}}} ,m + 1)} \!\!\!&\!\!\!  \cdots  \!\!\!&\!\!\! {g(k_{{\rm{D}}} ,k_{{\rm{D}}} )/\bar v _{k_{{\rm{D}}} } }  \\
\end{array}} \right]^{\text{T}}
\label{long_equation_1}
\end{split}
\end{eqnarray}
\hrule
\end{figure*}

\begin{itemize}
\item For $1 \leq m < k_{\text{D}}$, if $\mathbf{\bar P} \preceq \mathbf{P}(\mathbf{V}^{(m)}(\bar v_m)) \preceq
\mathbf{P}^{\text{max}}$, where $\mathbf{P}^{\text{max}}=\mathbf{P}=[p_1^{\text{max}}, p_2^{\text{max}}, \cdots, p_{k_{\text{D}}}^{\text{max}}]^{\text{T}}$, holds and $\mathbf{\bar P} \preceq
\mathbf{P}(\mathbf{V}^{(m)}(\bar v_{m+1})) \preceq
\mathbf{P}^{\text{max}}$ does not hold, the optimal power allocation in problem $(\text{P}_3)$ enables the optimal SINR vector $\mathbf{V}^{*}=[v_1^*, v_2^*, \cdots, v_{k_{\text{D}}}^*]$ to satisfy $\mathbf{V}^{*}=\mathbf{V}^{(m)}(v^{*})=[\underbrace
{v^{*}, \cdots, v^{*}}_{m}, \bar v_{m+1}, \cdots,
\bar v_{k_{\text{D}}}]$, where $\bar v_m \leq v^{*}< \bar
v_{m+1}$.

\item For $m = k_{\text{D}}$, if $\mathbf{\bar P} \preceq
\mathbf{P}(\mathbf{V}^{(k_{\text{D}} )}(\bar v_{k_{\text{D}}})) \preceq
\mathbf{P}^{\text{max}}$ holds, the optimal power allocation in problem
$(\text{P}_3)$ enables the optimal SINR vector $\mathbf{V}^{*}=[v_1^*, v_2^*, \cdots, v_{k_{\text{D}}}^*]$ to satisfy $\mathbf{V}^{*}= \mathbf{V}^{(m)}(v^{*})=[\underbrace
{v^{*}, \cdots, v^{*}}_{k_{\text{D}}}]$, where $v^{*}\geq \bar v_{k_{\text{D}}}$.
\end{itemize}

\begin{proof}
The proof is provided in Appendix C.
\end{proof}

Suppose that there exists a user $m$ ($1 \leq m <
k_{\text{D}}$), which satisfies $\mathbf{\bar P} \preceq
\mathbf{P}(\mathbf{V}^{(m)}(\bar v_m))\preceq
\mathbf{P}^{\text{max}}$ but does not satisfy $\mathbf{\bar P} \preceq
\mathbf{P}(\mathbf{V}^{(m)}(\bar v_{m+1})) \preceq
\mathbf{P}^{\text{max}}$. From Theorem 3, the optimal power allocation $\mathbf{P}^*$ in problem
$(\text{P}_3)$ enables $\mathbf{V}^{*}=[\underbrace
{v^{*}, \cdots, v^{*}}_{m}, \bar v_{m+1}, \cdots,
v_{k_{\text{D}}}]$ and can be written as
\begin{equation}
\mathbf{P}^*=\emph{H}^{-1}(\mathbf{V}^{*})\mathbf{N}, \label{P_star_D}
\end{equation}
which means that the optimal power allocation $\mathbf{P}^*$ can be
obtained by calculating $\emph{H}^{-1}(\mathbf{V}^{*})\mathbf{N}$.
However, the optimal SINR vector $\mathbf{V}^{*}$ is unknown to the BS. This makes it difficult to obtain $\mathbf{P}^*$ directly. Alternatively,
we develop a binary-search based algorithm to approach $ \mathbf{V}^{*}$ and
then obtain $\mathbf{P}^*$. Specifically, the optimal SINR vector is
$\mathbf{V}^{*}=\mathbf{V}^{(m)}(v^*)=[\underbrace {v^*, \cdots,
v^*}_{m}, \bar v_{m+1}, \cdots, v_{k_{\text{D}}}]$, where
 $\bar v_m \leq v^* < \bar v_{m+1}$. Consider that each transmit power in $\mathbf{P}^*$ increases if any SINR in $\mathbf{V}^{*}$ increases from Theorem 1. Then, we may approach
$\mathbf{V}^{*}$ by applying binary search between $\mathbf{V}^{(m)}(\bar
v_m)$ and $\mathbf{V}^{(m)}(\bar v_{m+1})$ subject to the transmit
power constraints.

Similarly, if $\mathbf{\bar P} \preceq
\emph{H}^{-1}(\mathbf{V}^{(k_{\text{D}})}(\bar V_{k_{\text{D}}}))\mathbf{N} \preceq
\mathbf{P}^{\text{max}}$ holds, the optimal SINRs can be denoted as $\mathbf{V}^*=\mathbf{V}^{(k_{\text{D}})}(v_{k_{\text{D}}})=[\underbrace {v^*,
\cdots, v^*}_{k_{\text{D}}}]$, where $v^* \geq \bar
v_{k_{\text{D}}}$. Meanwhile, consider that the optimal SINR vector
$v^*$ is upper-bounded by $\underset{m\in
\mathcal{S}^*_{\text{D}}}{\min} \ \frac{p_m^{\max}g(m, m)}{N_{m}}$, we
have $\bar v_{k_{\text{D}}} \leq v^* \le \underset{m\in
\mathcal{S}^*_{\text{D}}}{\min} \ \frac{p_m^{\max}g(m, m)}{N_{m}}$. Then, we may approach $\mathbf{V}^*$ by
applying binary search between $\mathbf{V}^{(k_{\text{D}})}(\bar
v_{k_{\text{D}}})$ and
$\mathbf{V}^{(k_{\text{D}})}\left(\underset{m\in
\mathcal{S}^*_{\text{D}}}{\min} \ \frac{p_m^{\max}g(m, m)}{N_{m}}\right)$
 subject to transmit power constraints. To summarize, we illustrate the binary-search based power allocation in \textbf{Algorithm \ref{NPAA}}.

\begin{algorithm}[htb]

\algsetup{linenosize=\footnotesize }

\caption{Binary-search based Power Allocation.}

\label{NPAA}

\begin{algorithmic}[1]

\REQUIRE ~~\ \

$m=1$, maximum tolerance error: $\epsilon_m$;

\ENSURE ~~\ \

\WHILE{$\mathbf{\bar P} \preceq \mathbf{P}(\mathbf{V}^{(m)}(\bar v_{m+1}))\preceq
\mathbf{P}^{\text{max}}$}

\STATE $m=m+1$;

\IF{$m \geq k_{\text{D}}$}

\STATE Break;

\ENDIF

\ENDWHILE

\IF{$m<k_{\text{D}}$}

\STATE $v_{\min}=\bar v_m$, $v_{\max}=\bar
v_{m+1}$, $v_{\text{mid}}=v_{\max}$;

\ELSE

\STATE  $v_{\min}=\bar v_{k_{\text{D}}}$,
$v_{\max}=\underset{n\in \mathcal{S}^*_{\text{D}}}{\min} \
\frac{p_n^{\max}g(n, n)}{N_{n}}$,
$v_{\text{mid}}=v_{\max}$;

\ENDIF

\WHILE{$\mathbf{\bar P} \preceq \mathbf{P}(\mathbf{V}^{(m)}(\bar v_{\text{mid}}))\preceq
\mathbf{P}^{\text{max}}$ does not hold or
$v_{\max}-v_{\min}>\epsilon_m$}

\STATE $v_{\text{mid}}=\frac{v_{\max}+v_{\min}}{2}$;

\IF{$\mathbf{\bar P} \preceq \mathbf{P}(\mathbf{V}^{(m)}(\bar v_{\text{mid}}))\preceq
\mathbf{P}^{\text{max}}$ does not hold}

\STATE $v_{\max}=v_{\text{mid}}$;

\ELSE

\STATE $v_{\min}=v_{\text{mid}}$;

\ENDIF

\ENDWHILE

\RETURN
$\mathbf{P}^*=\mathbf{P}(\mathbf{V}^{(m)}(v_{\text{mid}}))$;
\end{algorithmic}

\end{algorithm}

\subsection{Complexity Analysis}

The computational complexity of \textbf{Algorithm \ref{NPAA}} is dominated by two loops. In what follows, we will analyze the computational complexity of the two loop, respectively.

For the first loop from Line 1 to Line 6, we consider the worst case that there are $k_{\text{D}}$ rounds. In each round, the computational complexity is dominated by the calculation of  $\mathbf{P}(\Gamma^{(m)}(\bar v_{m+1}))=\emph{H}^{-1}( \mathbf{V}^{(m)}(\bar v_{m+1}))\mathbf{N}$, which consists of a
matrix inversion operation and a matrix multiplication operation. Since the
computational complexity of the inversion of a $k_{\text{D}} \times
k_{\text{D}}$ matrix $\emph{H}(\mathbf{V}^{(m)}(\bar v_{m+1}))$ is
$O(k_{\text{D}}^{2.373})$ and that of the multiplication of a
$k_{\text{D}} \times k_{\text{D}}$ matrix
$\emph{H}^{-1}(\mathbf{V}^{(m)}(\bar v_{m+1}))$ and a $k_{\text{D}} \times
1$ vector $\mathbf{N}$ is a $O(k_{\text{D}}^{2})$
\cite{Complexity_2}, we have the computational complexity to derive
$\mathbf{P}(\mathbf{V}^{(m)}(\bar v_{m+1}))$ is
$O(k_{\text{D}}^{2.373})+O(k_{\text{D}}^{2})=O(k_{\text{D}}^{2.373})$.
Then, the computational complexity of the first loop is $O(k_{\text{D}}^{3.373})$.

For the second loop from Line 12 to Line 19, the round of the binary search is
$O\left(\log\phi\right)$ \cite{Binary_search}, where $\phi=\frac{\underset{m\in
\mathcal{S}^*_{\text{D}}}{\min} \ \frac{p_m^{\max}g(m, m)}{N_{m}}-\bar
v_{k_{\text{D}}}}{e_m}$. In each round, the computational complexity is dominated by the calculation of $\mathbf{P}(\mathbf{V}^{(m)}(\bar v_{\text{mid}}))$ and thus is $O(k_{\text{D}}^{2.373})$. Then, the computational complexity of the second loop is $O( k_{\text{D}}^{2.373}\log\phi)$.

Consequently, the computational complexity of \textbf{Algorithm \ref{NPAA}} is $O(k_{\text{D}}^{3.373})+O( k_{\text{D}}^{2.373}\log\phi)$.

%
%From \cite{Convex}, the computational complexity of solving the a
%convex optimization problem \textbf{(P6)} with $k_{\text{DD}}$
%variables is $O(k_{\text{DD}}^{3.5})$.

%since the optimal power allocation $\mathbf{P}^*_i$ is
%non-decreasing in terms of the optimal SINR $\gamma^*_i$ from
%(\ref{optimal_equation}), i.e., $\mathbf{P}^{l+1}_i \ge
%\mathbf{P}^{l}_i$, there exists an $\mu$ $(0<\mu<1)$, such that
%\begin{equation}
%\mathbf{P}^{(l)}_i \ge \mu(\mathbf{P}^{(l+1)}_i).
%\label{iteration_complexity_1}
%\end{equation}
%
%Then, the number of iteration $L$ need to satisfy
%\begin{equation}
%\mathbf{P}^{(1)}_i \ge \mu \mathbf{P}^{(2)}_i \ge \cdots \ge
%\mu^{L-1}\mathbf{P}^{(L)}_i \ge \mu^L \mathbf{P}^{\text{max}}_u
%\bf{1},
%\end{equation}
%which means
%\begin{equation}
%\underset{m\subset \Omega_i}{\min} \  p_{i,m}^{(1)} \ge
%\mu^Lp_u^{\text{max}}, \ \forall, \ m \in \Omega_i.
%\end{equation}
%
%Thus, we have
%\begin{equation}
%L \ge \frac{\log \frac{p_u^{\text{max}}}{\underset{m\subset
%\Omega_i}{\min} \ p_{i,m}^{(1)}}}{\log\frac{1}{\mu}}.
%\end{equation}

\section{Numerical Results}
In this section, we will show the performance of our proposed
algorithms from numerical sides. To show the advantages of the proposed algorithms, we will also give the
performance comparison with the existing similar algorithms in \cite{D2D3}, \cite{D2D4}, \cite{DCPC}, and \cite{ITLinQ}.

\begin{figure}[!tp]
            \centering
            \includegraphics[scale=0.5]{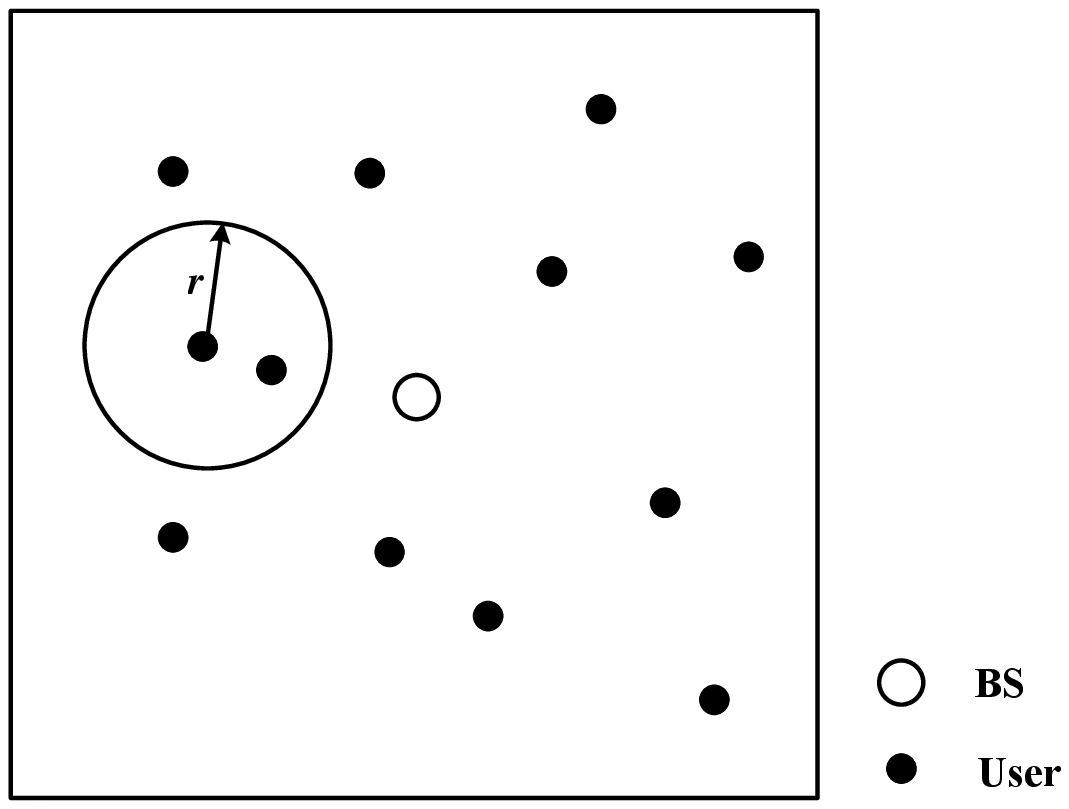}
            \caption{Simulation network model, where the empty circle denotes the BS and a solid circle denotes a user.}
            \label{Simulation_model}
 \end{figure}

In our simulation, we adopt the pathloss channel as $y=\frac{\lambda}{4\pi d}x+n_0$,
where $y$ is the received signal at the receiver, $x$ is the
transmit signal at the transmitter, $d$ is the distance between two
transceivers, $\lambda$ is the wavelength of the carrier, and $n_0$ is the AWGN with power spectral density $N_0=-170$ dBm/Hz at the
receiver. We assume that all the maximum transmit
power constraints are the same, i.e., $p^{\max}_m=20$
dBm, $\forall \ 1 \leq m \leq K$, and that the minimum acceptable SINRs at all the scheduled D2D receivers are the same, i.e., $ \check{v}_m=v_{\text{T}}, \ \forall \ m \in \mathcal{S}^*_{\text{D}}$, that the number of overall files is $N=1000$, and the bandwidth for D2D communication
is $1$ MHz. For fair comparison with \cite{D2D4}, we assume a square
cell with each side $1$ km as shown in Fig. \ref{Simulation_model}.

 \begin{figure}[!tp]
            \centering
            \includegraphics[scale=0.5]{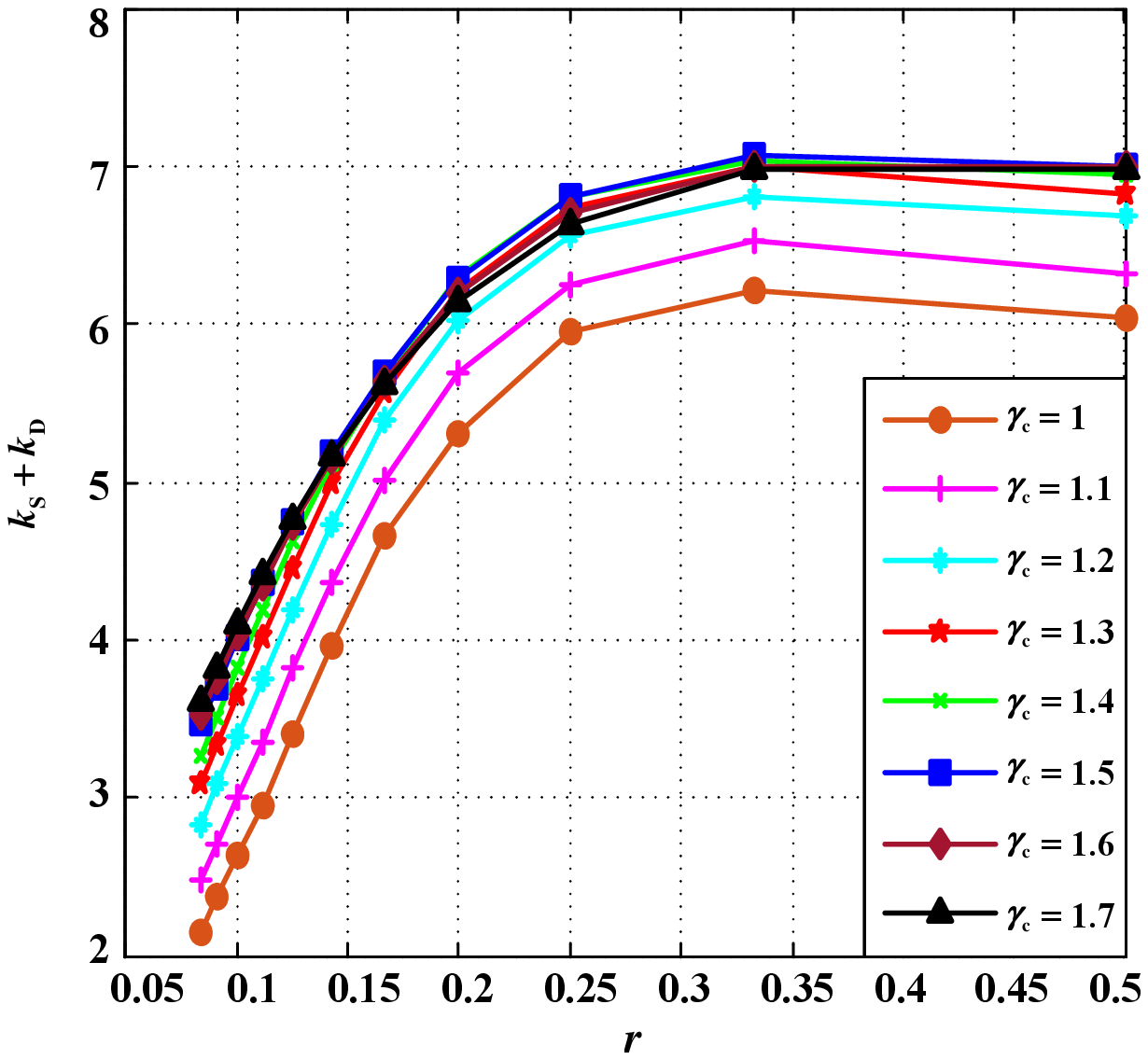}
            \caption{The number of users that can obtain the requested files from their own memory or from helpers through D2D transmissions, i.e., $k_{\text{S}}+k_{\text{D}}$, with different help distances $r$ and file caching coefficients $\gamma_c$, where $v_{\text{T}}=0$ dB, $c_{\text{s}}=0$ dB, $\gamma_r=0.6$, and $K=100$.}
            \label{Optimal_gammac_found}
 \end{figure}
Fig. \ref{Optimal_gammac_found} provides the number of users that can obtain the requested file from their own memory or from helpers through D2D transmissions, i.e., $k_{\text{S}}+k_{\text{D}}$, with different help distances $r$ and file caching coefficients $\gamma_c$. From this figure, the
$k_{\text{S}}+k_{\text{D}}$ first increases and then decreases
as $r$ grows. In fact, although $k_{\text{S}}$ is only determined by the caching coefficient, $k_{\text{D}}$ is directly affected by two factors. One is the number of the potential
D2D links $k_{\text{DB}}$. The other one is the interference among the scheduled D2D links. Then, $k_{\text{S}}+k_{\text{D}}$ is more sensitive to the variation of $k_{\text{D}}$ compared with $k_{\text{S}}$.  As $r$ grows, the chance that one user finds the requested file from
other users memory increases. This increases $k_{\text{DB}}$, $k_{\text{D}}$ and $k_{\text{S}}+k_{\text{D}}$. As $r$ continues to grow, the average distance between D2D transceivers
increases. This leads to higher transmit power at each D2D transmitter. Thus, the mutual interference among the scheduled D2D links increases,
and decreases $k_{\text{S}}+k_{\text{D}}$. Besides, we
observe that the optimal $r$ and $\gamma_c$ in terms of the largest $k_{\text{S}}+k_{\text{D}}$ is about $0.33$ km and $1.5$, respectively.

 \begin{figure}[!tp]
            \centering
            \includegraphics[scale=0.5]{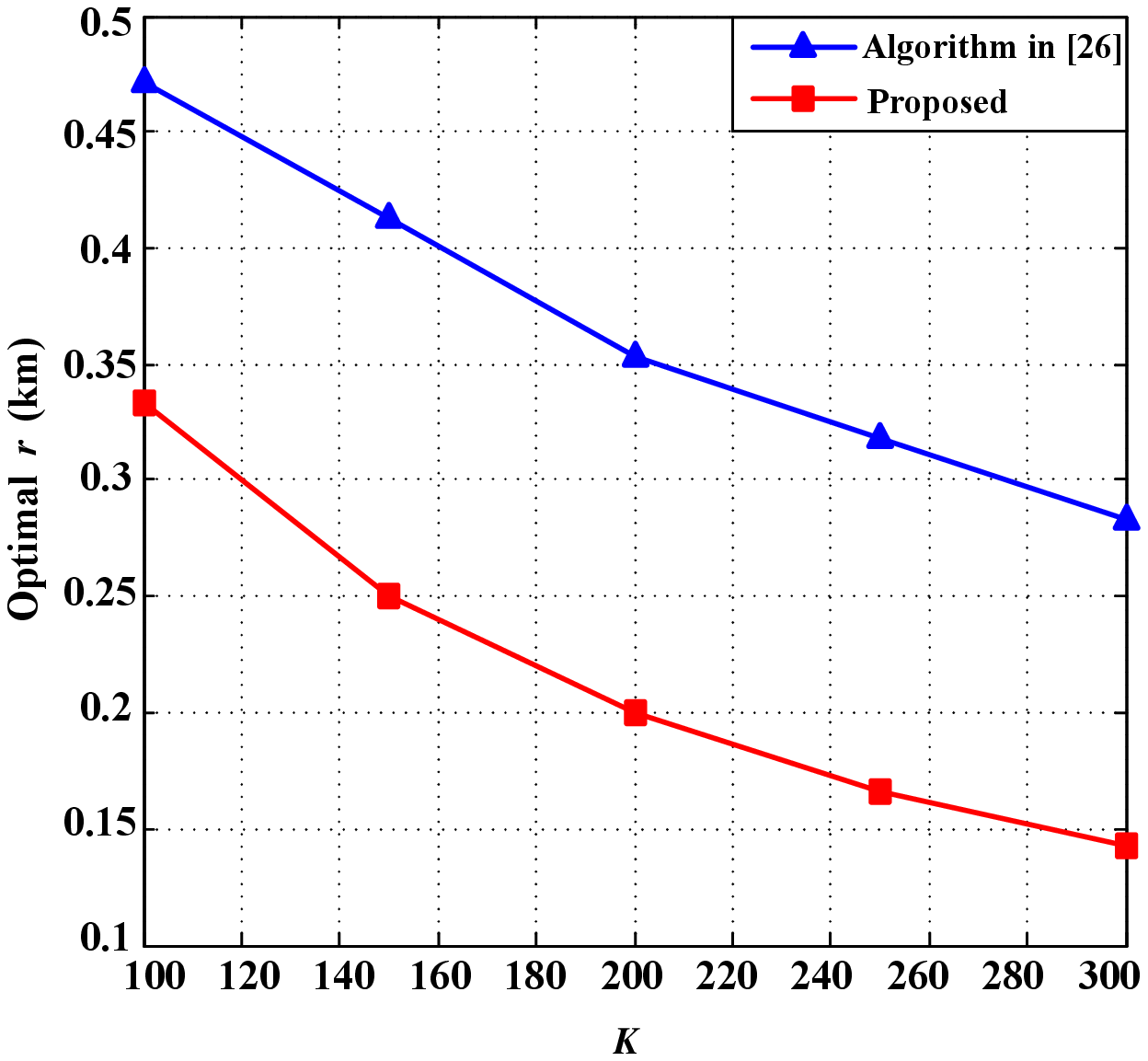}
            \caption{The Optimal help distance versus different number of users in the cell, where $v_{\text{T}}=0$ dB, $c_{\text{s}}=0$ dB, $\gamma_r=0.6$.}
            \label{Optimal_gammar}
 \end{figure}

Fig. \ref{Optimal_gammar} gives the optimal $r$ with different
number of users $K$. For comparison, we
also provide the optimal $r$ in \cite{D2D4}. It is clear that the optimal $r$ decreases as
$K$ grows. For small $K$, the number of the potential D2D links $k_{\text{DB}}$ limits the number of users that can obtain the requested file from their own memory or from helpers through D2D transmissions, i.e., $k_{\text{S}}+k_{\text{D}}$. To maximize $k_{\text{S}}+k_{\text{D}}$, the optimal
$r$ needs to be large to increase the potential D2D
links. As $K$ grows, a small $r$ may result
in a large number of the potential D2D links $k_{\text{DB}}$, which generates strong
mutual interference to each other and limits the number of the
scheduled D2D links. Thus, a smaller optimal $r$ is needed for a
larger $K$. Besides, we observe that the optimal $r$ in our algorithm is smaller than the optimal $r$ in \cite{D2D4}. This is reasonable since each user in our algorithm can establish more potential D2D links than the algorithm in \cite{D2D4} for a given number of users. Then, a smaller optimal $r$ is needed to maximize $k_{\text{S}}+k_{\text{D}}$.

 \begin{figure}[!tp]
            \centering
            \includegraphics[scale=0.5]{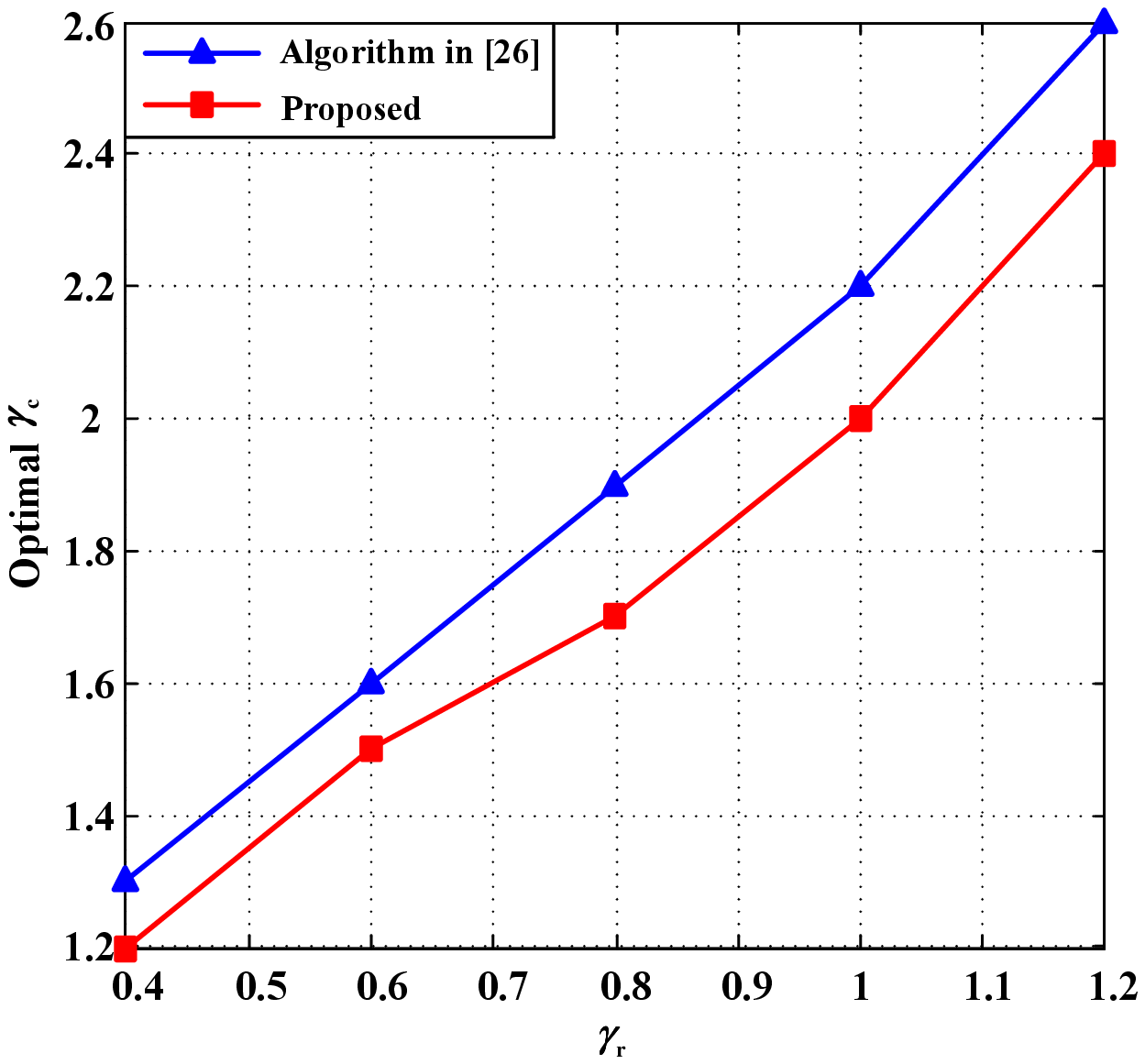}
            \caption{The Optimal file caching coefficient versus different file request coefficients, where $v_{\text{T}}=0$ dB, $c_{\text{s}}=0$ dB, $K=100$.}
            \label{Optimal_gammac}
 \end{figure}

Fig. \ref{Optimal_gammac} gives the optimal $\gamma_c$ with
different file request coefficients $\gamma_r$. For comparison, we
also provide the optimal $\gamma_c$ in \cite{D2D4}. It is observed that, the optimal $r$ increases as $\gamma_c$ grows. When
$\gamma_r$ is small, more different files are requested by the users
in the cell. Then, users need to cache more different files to
maximize the number of the scheduled D2D links. Thus, the optimal $\gamma_c$ is small.
Otherwise, a big optimal $\gamma_c$ is required. Besides, we observe that the
optimal $\gamma_c$ in our algorithm is smaller than the optimal $\gamma_c$ in \cite{D2D4}. This is because each user in our algorithm can potentially help more users than that in \cite{D2D4}. Then, users in our algorithm is required to cache more different files for D2D transmissions. Thus, a smaller optimal $\gamma_c$ is needed in our algorithm.

 \begin{figure}[!tp]
            \centering
            \includegraphics[scale=0.5]{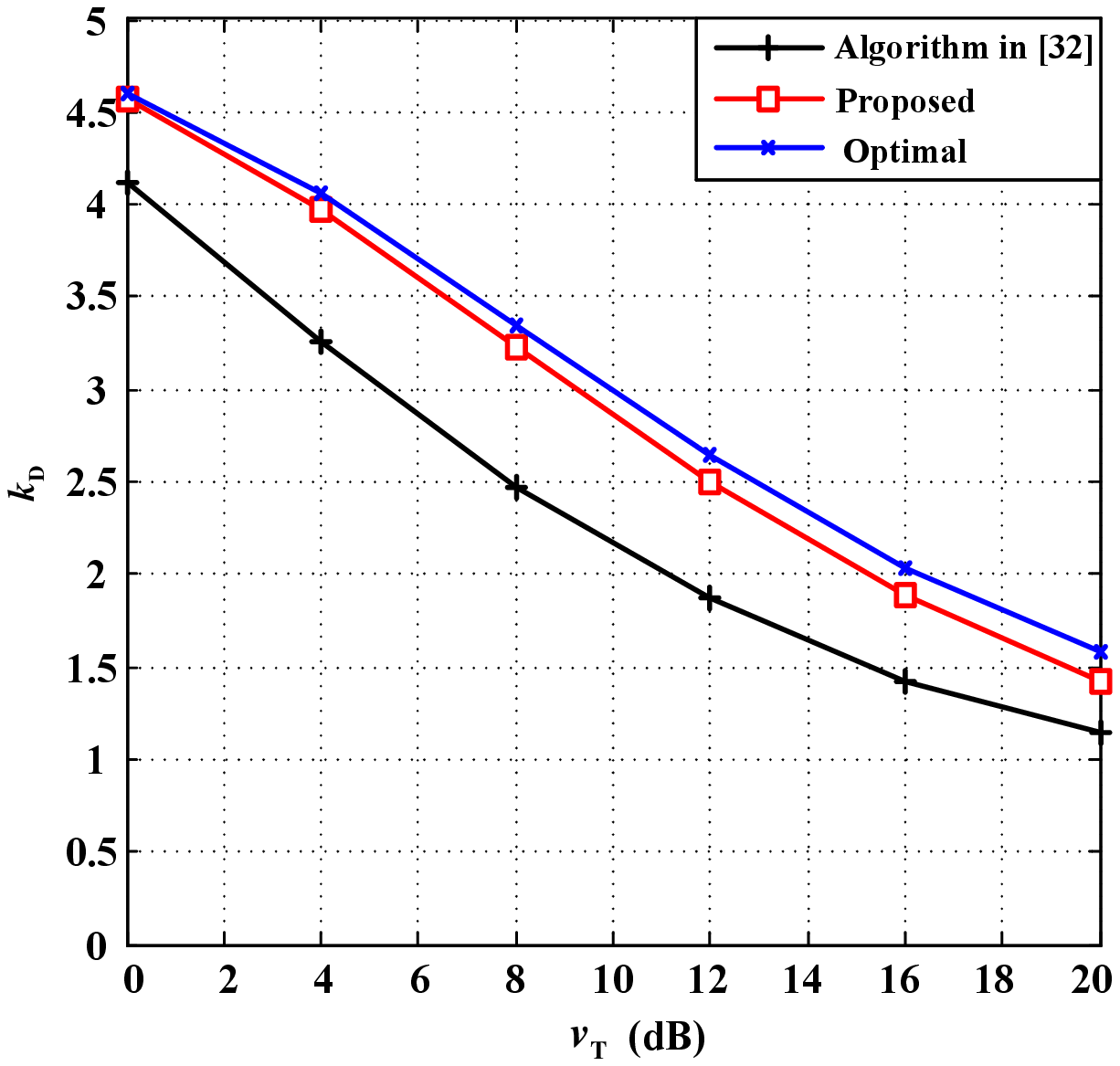}
            \caption{Comparison of the numbers of the scheduled D2D links with different scheduling algorithms, where $v_{\text{T}}=0$ dB, $c_{\text{s}}=0$ dB, $\gamma_r=0.6$, $\gamma_c=1.5$, $r=\frac{1}{7}$ km.}
            \label{Compare_with_optimal}
 \end{figure}

Fig. \ref{Compare_with_optimal} compares the number of the scheduled
D2D links of the proposed CPC-based D2D link scheduling algorithm with the DCPC-based D2D link scheduling
algorithm in \cite{DCPC} and the optimal schedule algorithm from
the exhaustive search. Since the computational complexity of the
exhaustive search with a large number of potential D2D links $k_{\text{DB}}$ is too
high, we limit the number of the potential D2D links by choosing a
small help distance instead of the optimal one. It could be concluded from this figure that,
the performance of the proposed algorithm is between the performance of the
optimal schedule algorithm and the DCPC-based schedule algorithm in
\cite{DCPC}. This indicates that the CPC-based D2D link scheduling algorithm is more flexible than that the DCPC-based D2D link scheduling algorithm in \cite{DCPC}.

 \begin{figure}[!tp]
            \centering
            \includegraphics[scale=0.5]{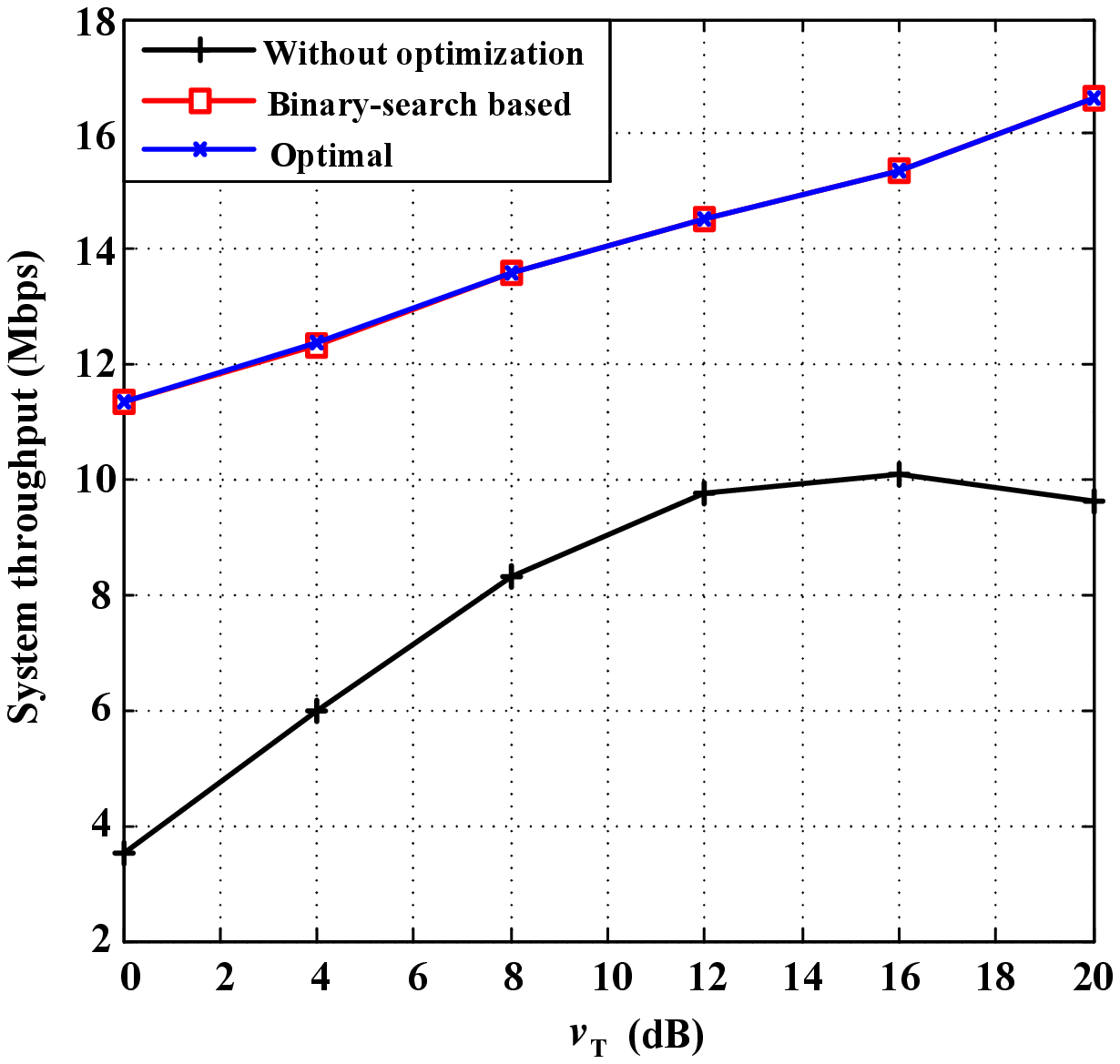}
            \caption{Comparison of the system throughput with different power allocation algorithms, where $c_{\text{s}}=v_{\text{T}}$, $K=100$, $\gamma_r=0.6$, $\gamma_c=1.5$, and $r=\frac{1}{7}$ km. }
            \label{Transmission_rate}
 \end{figure}

Fig. \ref{Transmission_rate} shows the system throughput of the
scheduled D2D links with the binary-search based power
allocation algorithm in \textbf{Algorithm \ref{NPAA}}. Here, the system throughput is calculated by
$\underset{m\in \mathcal{S}^*_{\text{D}}}{\sum} \log(v_m+1)$. For
comparison, we provide the performance of the optimal power
allocation algorithm, where a convex optimization toolbox is adopted,
and the algorithm without optimization, where each D2D
link works with the minimum SINR. From Fig. \ref{Transmission_rate}, the performance curve of the proposed
binary-search based power allocation algorithm almost overlaps with
the optimal one. This validates our analysis and indicates that our
proposed algorithm may achieve similar rate performance with the
optimal algorithm and outperforms the algorithm without optimized
power allocation, i.e., more than $40\%$ improvement. It should be noted that the proposed optimal power
allocation algorithm is based on max-min fairness, which achieves perfect fairness. In fact, the
system throughput can be further improved if some unfairness can be
tolerated \cite{DCPC}.

\begin{figure}[!tp]
            \centering
            \includegraphics[scale=0.5]{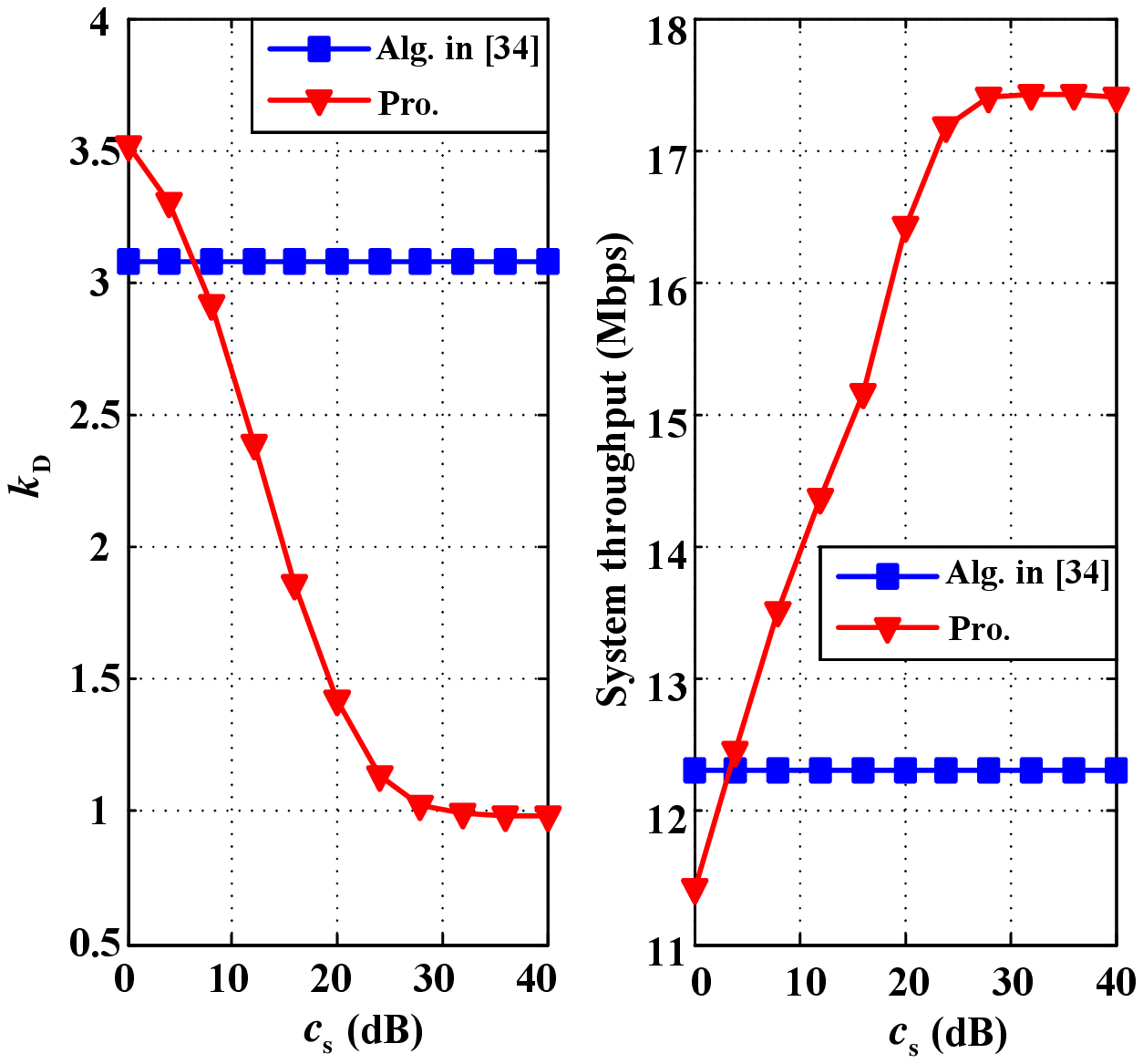}
            \caption{Performance comparison with scheduling algorithm in \cite{ITLinQ} for different scheduling coefficient $c_{\text{s}}$, where $v_{\text{T}}=0$ dB, $K=100$, $\gamma_c=1.5$ and $r=\frac{1}{7}$ km. }
            \label{Compared_with_ITinQ_1}
 \end{figure}

Fig. \ref{Compared_with_ITinQ_1} shows the number of the D2D links and the
system throughput with different scheduling coefficient
$c_{\text{s}}$. We observe that the
number of the D2D links decreases as $c_{\text{s}}$ grows. This coincides with the intuition
that a larger $c_{\text{s}}$ means a stricter condition on the scheduled
D2D links. Then, the D2D links with stronger communication channels
and weaker interference channels are scheduled. On the other hand,
the system throughput increases as $c_{\text{s}}$ grows. This is
quite reasonable since fewer D2D links mean less mutual interference,
which boosts the system throughput. On the other hand, the system throughput of the optimal scheduling in terms of the largest number of the scheduled D2D links is much smaller than the
largest system throughput. Thus, the number of the scheduled D2D
links and the system throughput can be balanced by choosing a proper
$c_{\text{s}}$.

Meanwhile, we provide the performance of the algorithm in
\cite{ITLinQ}, where only the information theoretic independent sets
are scheduled. Note that there is no power allocation in
\cite{ITLinQ}, we apply \textbf{Algorithm \ref{NPAA}}
in \cite{ITLinQ} for fair comparison. From the figure, the number of
the D2D links and the system throughput are constant since the
scheduling algorithm in \cite{ITLinQ} is not affected by
$c_{\text{s}}$. We also observe that the number of the scheduled D2D
links and the system throughput with the proposed algorithms can
simultaneously outperform those with the algorithm in \cite{ITLinQ}.
For instance, $c_{\text{s}}$ should be chosen around between
four and six in this figure, e.g., $c_{\text{s}}=5$ dB. In this way,
we may choose proper values of $c_{\text{s}}$ for different system
parameters.

 \begin{figure}[!tp]
            \centering
            \includegraphics[scale=0.5]{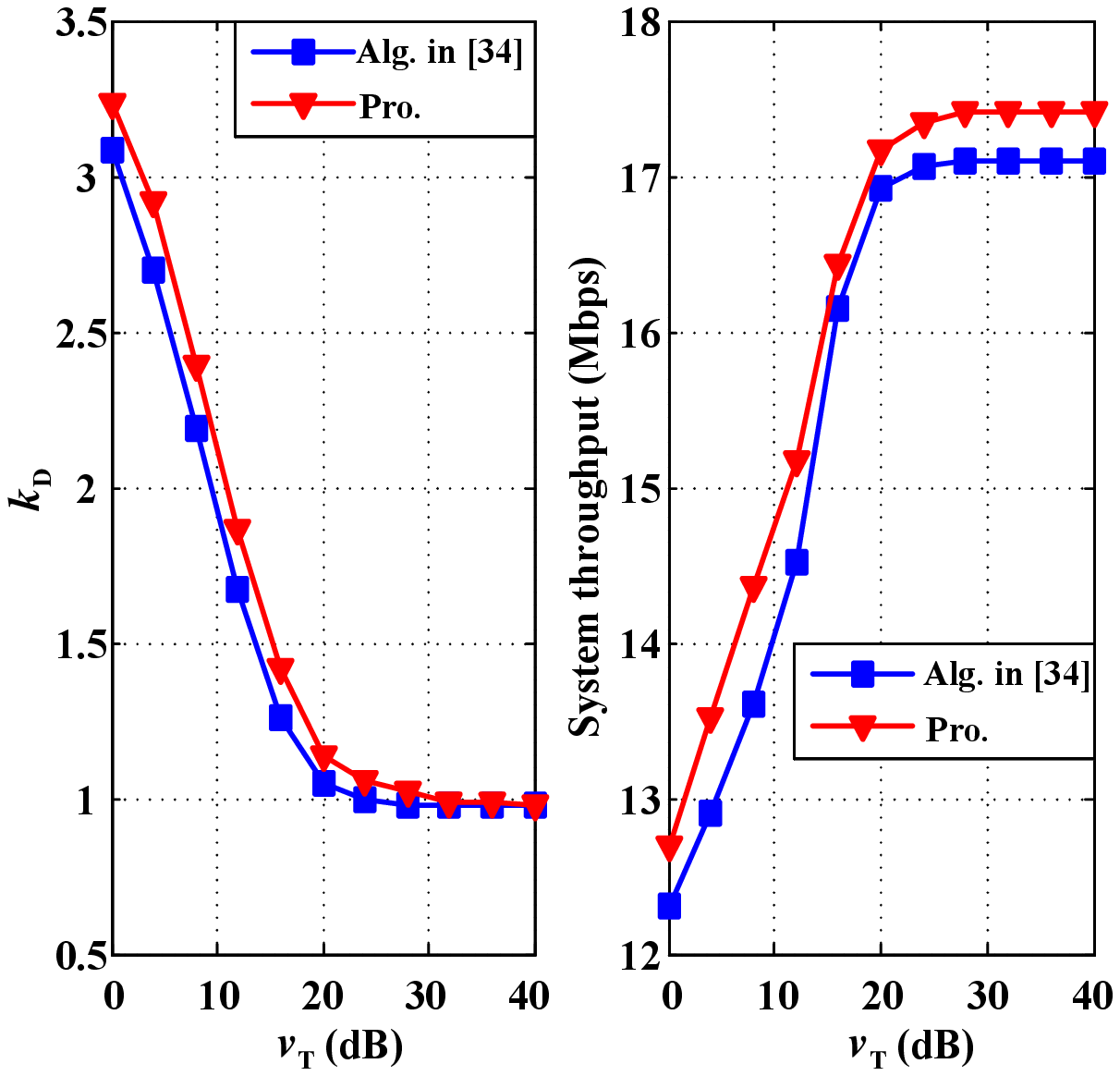}
            \caption{Performance comparison with the scheduling algorithm in \cite{ITLinQ} in terms of $k_{\text{D}}$ and system throughput for different $v_{\text{T}}$, where $K=100$,  $\gamma_c=1.5$ and $r=\frac{1}{7}$ km.}
            \label{Compared_with_ITinQ_2}
 \end{figure}

\begin{table}[!hbp] 
\centering
\begin{tabular}{|c|c|c|c|c|c|c|c|c|c|c|c|}
  \hline
  $v_{\text{T}}$ (dB)& 0 & 4 & 8 & 12 & 16 & 20 & 24 & 28 & 32 & 36 & 40 \\
  \hline
  $c_{\text{s}}$ (dB) & 5 & 8 & 12 & 16 & 20 & 24 & 26 & 28 & 32 & 36 & 40 \\
  \hline
\end{tabular}
\caption{The values of $c_{\text{s}}$ with different $v_{\text{T}}$ in Fig. \ref{Compared_with_ITinQ_2}. }
\end{table}

Fig. \ref{Compared_with_ITinQ_2} compares the proposed algorithm with the algorithms in \cite{ITLinQ} in terms of the number of the scheduled D2D links and the system throughput. With the same selection method of $c_{\text{s}}$ in Fig. \ref{Compared_with_ITinQ_1}, we choose the values of $c_{\text{s}}$ with different $v_{\text{T}}$ in Tab. I. It is observed that the proposed algorithms outperform the algorithm in \cite{ITLinQ} in terms of both the number of the scheduled D2D links and the system throughput.

 \begin{figure}[!tp]
            \centering
            \includegraphics[scale=0.5]{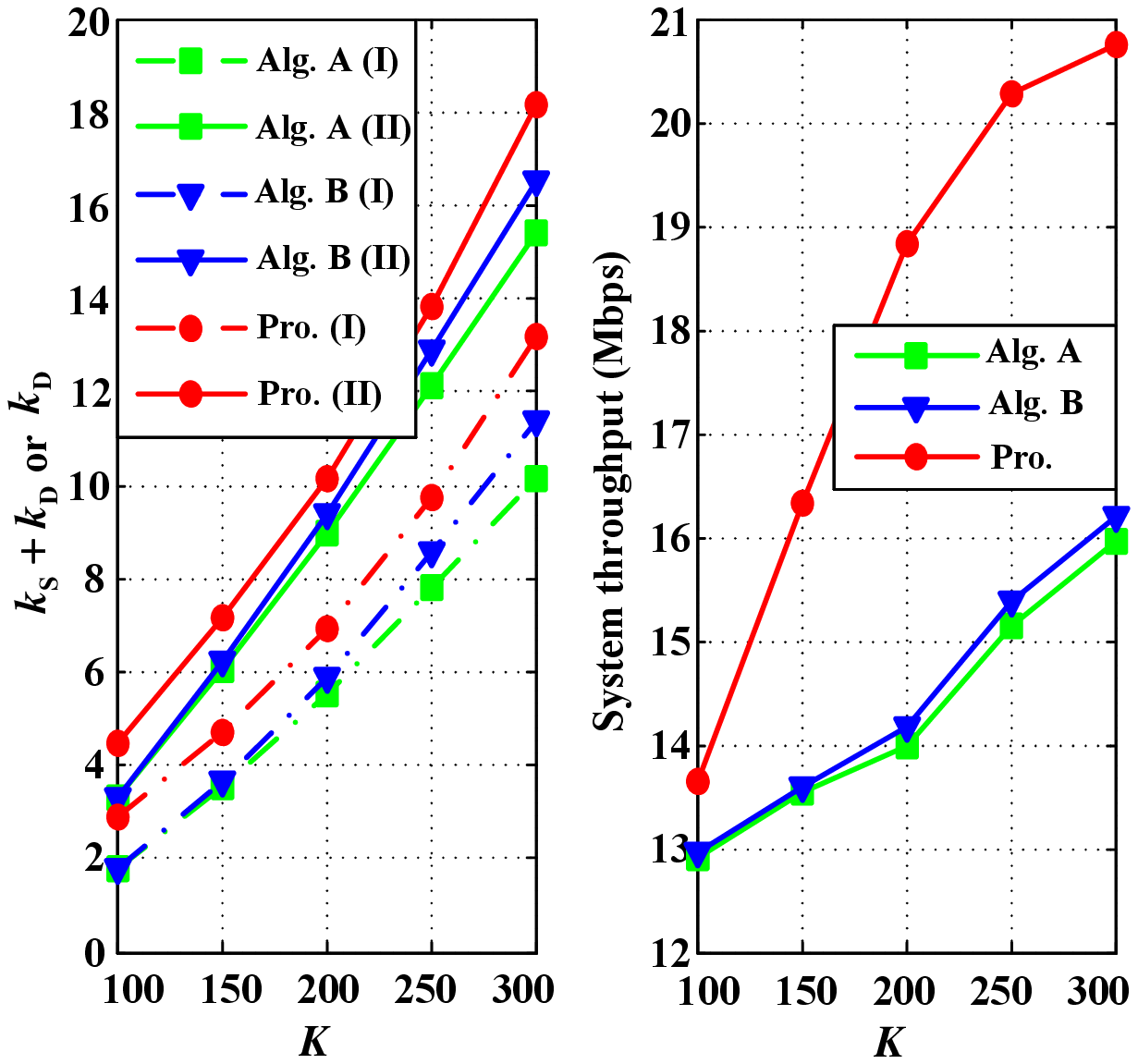}
            \caption{Performance comparison with the scheduling algorithms in \cite{D2D3} and \cite{D2D4} in terms of $k_{\text{S}}+k_{\text{D}}$, $k_{\text{D}}$, and system throughput for different number of users in the cell, where $v_{\text{T}}=0$ dB, $\gamma_r=0.6$, $\gamma_c=1.5$, $r=\frac{1}{7}$ km. }
            \label{Compared_with_Cluster}
 \end{figure}

Fig. \ref{Compared_with_Cluster} compares the proposed algorithm
(Pro. in the figure) with algorithm A (Alg. A in the figure) and
algorithm B (Alg. B in the figure) in terms of the number of the
scheduled D2D links and system throughput. In this figure, (I) and
(II) denote $k_{\text{D}}$ and $k_{\text{S}}+k_{\text{D}}$,
respectively. In algorithm A, we adopt the Zipf-distribution caching and scheduling scheme in \cite{D2D4}, and the proposed optimal power allocation algorithm with perfect fairness
meanwhile considering the minimum acceptable SINR constraint, i.e.,
$v_{\text{T}}=0$ dB. In algorithm B, we adopt the optimal cluster-based caching and scheduling scheme in \cite{D2D3}, and the
proposed power allocation algorithm with perfect fairness meanwhile
considering the minimum acceptable SINR constraint, i.e.,
$v_{\text{T}}=0$ dB. The scheduling coefficient
$c_\text{s}$ of the proposed algorithm is obtained with the same selection method in Fig.
\ref{Compared_with_ITinQ_1} for each $K$. That is, $c_\text{s}$ is
chosen to be $10$ dB, $8$ dB, $6$ dB, $4$ dB, and $2$ dB when $K$ is
$100$, $150$, $200$, $250$, and $300$, respectively. From this
figure, the proposed algorithms outperform the algorithms in \cite{D2D3} and \cite{D2D4} in terms of the number of the scheduled D2D links, the number of users that can obtain the requested files either from their own memory or from helpers through D2D transmissions, and the system throughput. This is intuitive since the proposed algorithms create more D2D links with strong communication channels and weak interference channels. Then, the number of the scheduled D2D links and the system throughput can be simultaneously enhanced by efficiently scheduling and power allocation.

 \begin{figure}[!tp]
            \centering
            \includegraphics[scale=0.5]{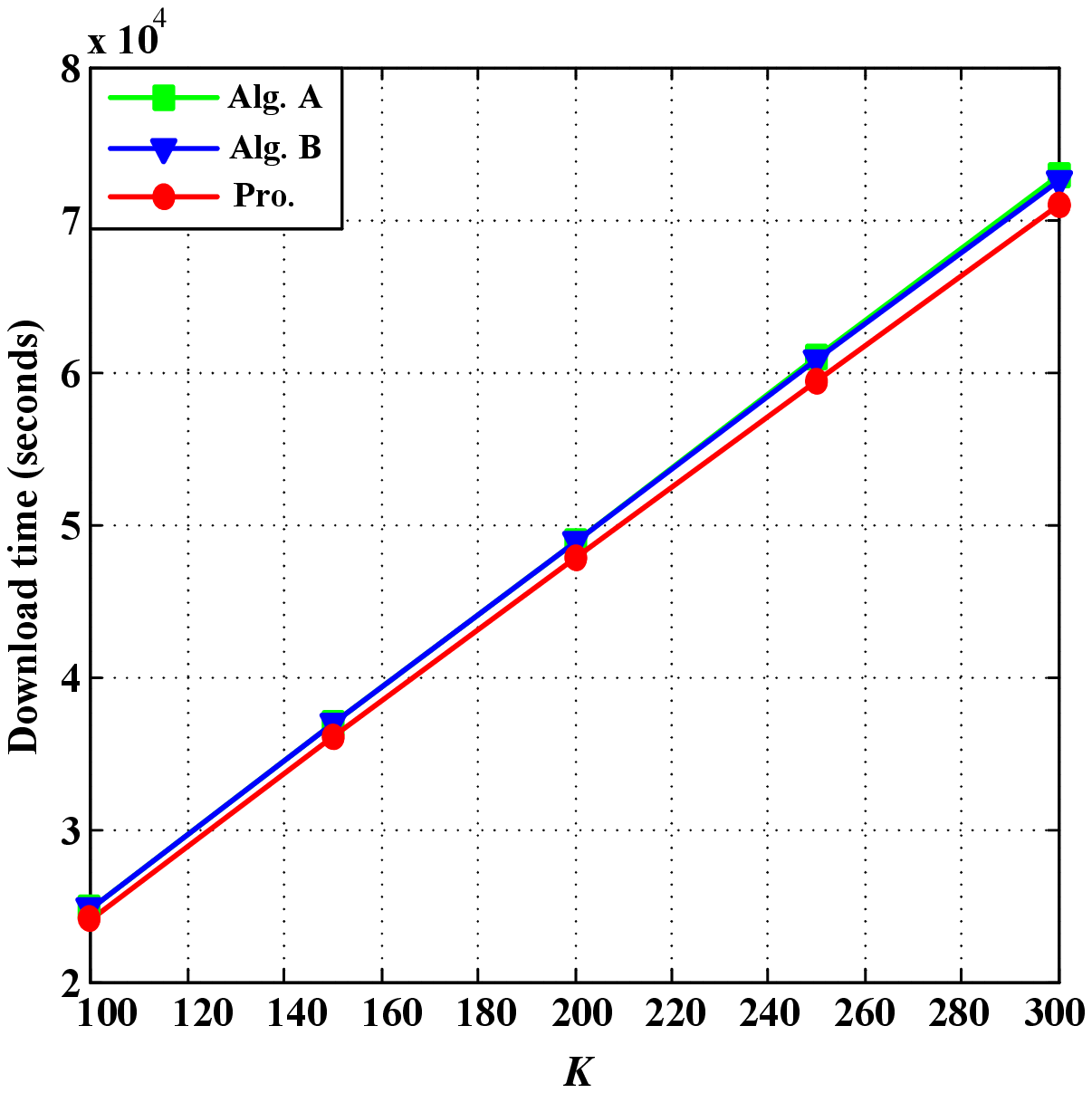}
            \caption{Performance comparison with the scheduling algorithm in \cite{D2D4} in terms of the download time for different number of users in the cell, where $v_{\text{T}}=0$ dB, $\gamma_r=0.6$, $\gamma_c=1.5$, $r=\frac{1}{7}$ km. }
            \label{Latency}
 \end{figure}

Fig. \ref{Latency} compares the three algorithms (Alg. A, Alg. B, and Pro. in Fig. \ref{Compared_with_Cluster}) in terms of the download time, which is calculated by $\sum_{m\in S_{\text{B}} \cup S_{\text{DB}} \backslash S^*_{\text{D}}}w^{\text{BS}}_m+\sum_{m\in S^*_{\text{D}}}w^{\text{D2D}}_m$ \cite{D2D4}, where $w^{\text{BS}}_m$ is the download time of a video file from the BS to user $m$ and $w^{\text{D2D}}_m$ is the download time of a video file from a helper to user $m$ through an one-hop D2D transmission. Specifically, $w^{\text{BS}}_m$ is calculated by $w^{\text{BS}}_m=\frac{L_{\text{video}}}{R^{\text{BS}}_m}$, where $L_{\text{video}}$ is the length of a file, say $L_{\text{video}}=30$ MB in the simulation, and $R^{\text{BS}}_m$ is the transmission rate from the BS to user $m$ and is assumed to be $R^{\text{BS}}_m=120$ kbps \cite{D2D4}, $\forall \ m\in S_{\text{B}} \cup S_{\text{DB}} \backslash S^*_{\text{D}}$. Similarly, $w^{\text{D2D}}_m$ is calculated by $w^{\text{D2D}}_m=\frac{L_{\text{video}}}{R^{\text{D2D}}_m}$, where $R^{\text{D2D}}_m$ is the transmission rate of D2D link $l_m$ and is assumed to be $R^{\text{D2D}}_m=\log(v_m+1)$ Mbps. From this figure, the proposed algorithms outperforms both Alg. A and Alg. B. This is reasonable since more users in the proposed algorithms can obtain the requested files from their own memory or from helpers through D2D transmissions compared with those in either Alg. A or Alg. B (as shown in Fig. \ref{Compared_with_Cluster}). Thus, the average download time is reduced with the proposed algorithms.

\section{Conclusions and Future Discussions}

We have studied the efficient scheduling and power allocation of D2D-assisted wireless caching networks. We formulate a joint D2D link scheduling and power allocation problem to maximize the system throughput. However, the problem is non-convex and obtaining the optimal solution is computationally hard. Alternatively, we seek to obtain a suboptimal solution with reasonable complexity. Briefly, we intend to schedule the D2D links with strong communication channels and weak interference channels and allocate the power to the scheduled D2D links fairly. Thus, we decompose the system throughput maximization problem into a D2D link scheduling problem and an optimal power allocation problem. To solve the two subproblems, we first develop a D2D link scheduling algorithm to increase the number of the scheduled D2D links satisfying both the SINR and the transmit power constraints. Then, we develop an optimal power allocation algorithm to maximize the minimum transmission rate of the scheduled D2D links. Numerical results indicate that both the number of the scheduled D2D links and the system throughput can be improved simultaneously with the Zipf-distribution caching scheme, the proposed D2D link scheduling algorithm, and the proposed optimal power allocation algorithm compared with the state of arts.

From the results in this paper, we also conclude that the D2D communication in a wireless caching network can be enhanced by creating more D2D links with strong communication channels and weak interference channels. This can be achieved by optimized caching, the proposed D2D links scheduling algorithm, and the proposed optimal power allocation algorithm. Although the optimized Zipf-distribution caching scheme in this paper is not generally optimal, the proposed D2D link scheduling algorithm and the proposed optimal power allocation algorithm can be used for any caching scheme. Ideally, the joint caching, scheduling, and power allocation algorithm should be able to further improve D2D communication. However, even the optimal caching distribution of the general model in our paper is quite hard. A further study on the optimal caching scheme is beyond the scope of this paper.

Throughout this paper, we assume one channel to simplify the illustration of our algorithms. In fact, the proposed D2D link scheduling algorithm and the proposed optimal power allocation algorithm can be used in multi-channel systems, e.g., FDMA and TDMA. More specifically, D2D links are iteratively scheduled in a unique channel, i.e., a frequency band (corresponding to FDMA) or a time slot (corresponding to TDMA). This process terminates in two cases. One is that there is no more channels to accommodate the scheduled D2D links. The other one is that all the D2D links have been scheduled. Since more wireless resource is used in multi-channel systems, a better system performance is expected.

Besides, both the proposed D2D link scheduling algorithm and the proposed optimal power allocation algorithm are achieved in a centralized manner. That is, all the calculations are conducted at the BS. This requires the BS to keep track of the cached content in each memory and collect the channel state information among different potential D2D links. One direction of the future work is to develop decentralized algorithms of scheduling and power allocation in the D2D assisted wireless caching networks.

\section{Appendix}

\subsection{Proof of Theorem 1}
To prove Theorem 1, we will first prove part I: each transmit power $p_n$, $\forall \ n \in \mathcal{S}_{\text{D}}$, increases if any SINR $v_m$, $ m \in \mathcal{S}_{\text{D}}$, in $(\text{P}_1)$ increases. Then, we will prove part II: the value of $|\mathcal{S}_{\text{D}}|$ remains constant or decreases if any SINR $v_m$, $ m \in \mathcal{S}_{\text{D}}$, in $(\text{P}_1)$ increases.
\subsubsection{Proof of part I in Theorem 1}
Suppose that there is a power
increment $\Delta p_m$ and a SINR increment $\Delta v_m$ for D2D link $l_m$ in $\mathcal{S}_{\text{D}}$ such that
 \begin{equation}
v_{m}+\Delta v_m=\frac{( p_{m}+\Delta p_m) g(m,
m)}{\sum_{n\in \mathcal{S}_{\text{D}}, n\neq m} p_{n} g(n,
m)+N_{m}}. \label{gamma_m1}
\end{equation}
Then, there must be a SINR decrement $\Delta v_n$ for any other D2D link $l_n$ ($\forall
n\ \in \mathcal{S}_{\text{D}}, n \neq m$) such that
\begin{equation}
%\begin{split}
v_{n}-\Delta v_n=
\frac{p_{n} g(n, n)}{\sum_{k\in
\mathcal{S}_{\text{D}}, k\neq n, k\neq m}  p_{k} g(k, n)+(
p_m+\Delta p_m)g(m,n)+N_{n}}. \label{gamma_m2}
%\end{split}
\end{equation}
To remain the SINR $v_{n}$ at D2D link $l_n$, $ p_n$
($\forall \ n\in \mathcal{S}_{\text{D}}, n \neq m$) will be increased by
$\Delta p_n$, such that
\begin{equation}
v_{n}=\frac{( p_{n}+\Delta p_n) g(n, n)}{\sum_{k\in
\mathcal{S}_{\text{D}}, k\neq n} (p_{k}+\Delta p_k) g(k, n)+N_{n}}.
\label{gamma_m3}
\end{equation}
Thus, each $ p_n$, $n \in \mathcal{S}_{\text{D}}$ increases if there is an increment of
any $v_m$, $ m \in \mathcal{S}_{\text{D}}$.

\subsubsection{Proof of part II in Theorem 1}

Suppose that the D2D links in $\mathcal{L}_{\text{D}}$ satisfy all the
constraints in problem $(\text{P}_1)$ with transmit power
$\mathbf{P}=\{ p_1, p_2, \cdots,
p_{k_{\text{D}}}\}$. If there is a SINR increment $\Delta v_m$
at D2D link $l_m$, the transmit power should be increased to
$\mathbf{P}+\Delta \mathbf{P}=\{ p_1+\Delta p_1,
p_2+\Delta p_2, \cdots, p_{k_{\text{D}}}+\Delta
p_{k_{\text{D}}}\}$ to satisfy all the minimum acceptable SINRs. If $p_n+\Delta p_n \leq p_n^{\max}$, $\forall \ n \in \mathcal{S}_{\text{D}}$, all the D2D links in $\mathcal{L}_{\text{D}}$ can still be satisfied with the minimum SINR constraints and number of the scheduled D2D
links remains. If there exists a
$\Delta p_n$ ($n\in \mathcal{S}_{\text{D}}$) satisfying $\bar p_n+\Delta p_n>p_n^{\max}$ and
\begin{equation}
v_{n}=\frac{p_n^{\max} g(n, n)}{\sum_{k\in \mathcal{S}_{\text{D}},
k\neq n} (p_{k}+\Delta p_k) g(k, n)+N_{n}} < \bar v_n.
\label{gamma_m4}
\end{equation}
Then, the D2D link $l_n$ cannot be satisfied with the minimum
SINR constraint. This reduces the number of the scheduled D2D
transmissions. This completes the proof of Theorem 1.

\subsection{Proof of Theorem 2}
The regular admission control problem with QoS constraint has been
proved to be NP-hard in \cite{DCPC} and is a special case (Case II)
of problem $(\text{P}_2)$. Specifically, the problem
$(\text{P}_2)$ reduces to a regular admission control problem
when any two D2D links in $\mathcal{L}_{\text{D}}$ do not share
the same users. Thus, problem $(\text{P}_2)$ is also NP-hard.

\subsection{Proof of Theorem 3}
In this part, we will prove Theorem 3. Firstly, we will give the
proof of the first case that there is a user $m$ ($0<m <
k_{\text{D}}$) satisfying $\mathbf{\bar P} \preceq
\mathbf{P}(\mathbf{V}^{(m)}(\bar v_m)) \preceq
\mathbf{P}^{\text{max}}$ but not satisfying $\mathbf{\bar P} \preceq
\mathbf{P}(\mathbf{V}^{(m)}(\bar v_{m+1})) \preceq
\mathbf{P}^{\text{max}}$, and prove that the optimal SINRs of the
scheduled D2D links is $\mathbf{V}^{*}=[\underbrace {v^*, \cdots,
v^*}_{m}, \bar v_{m+1}, \cdots, \bar v_{k_{\text{D}}}]$,
where $ \bar v_m \leq v^* \leq \bar v_{m+1}$. Then,
we will provide the proof of the second second case that if
$\mathbf{\bar P} \preceq \mathbf{P}(\mathbf{V}^{(k_{\text{D}})}(\bar v_{k_{\text{D}}}))
\preceq \mathbf{P}^{\text{max}}$ holds, the optimal SINRs of the
scheduled D2D links is $\mathbf{V}^{*}=[\underbrace {v^*, \cdots,
v^*}_{k_{\text{D}}}]$,
where $ v^* \geq \bar v_{k_{\text{D}}}$.

\subsubsection{Proof of the First Case}
We will prove this by two steps by contradiction, the first step is
to prove the first $m$ optimal SINRs are identical, i.e.,
$v_n^*=v^*$ for $1 \leq n \leq m$. The other step is to
prove the last $k_{\text{D}}-m$ optimal SINRs are the minimum SINR
constraints, i.e., $v_n^*=\bar v_n$ for $ m+1 \leq n \leq
k_{\text{D}}$.

Denote the optimal SINRs of the scheduled D2D links as
$\mathbf{V}^{*}=[\underbrace {v_1^*, \cdots, v_m^*}_{m},
v_{m+1}^*, \cdots, v_{k_{\text{D}}}^*]$. Since
$\mathbf{\bar P} \preceq \mathbf{P}(\mathbf{V}^{(m)}(\bar v_m)) \preceq
\mathbf{P}^{\text{max}}$ holds and $\mathbf{\bar P} \preceq
\mathbf{P}(\mathbf{V}^{(m)}(\bar v_{m+1})) \preceq
\mathbf{P}^{\text{max}}$ does not hold, we have $\bar v_{m}
\leq v_n^* < \bar v_{m+1}$ for $1\leq n \leq m$ and
$v_n^* \geq \bar v_n$ for $m+1 \leq n \leq k_{\text{D}}$.
If we denote the optimal power allocation as $\mathbf{P}^*=[p_1^*,
p_2^*, \cdots, p_{k_{\text{D}}}^* ]$, we have
\begin{equation}
\frac{p_{n}^*g(n, n)}{\sum_{k =1, k \neq n}^{k_{\text{D}}}p_{
k}^*g(k, n)+N_n}=v_{n}^* \geq \bar v_m \geq \bar v_n,
\forall \ 1\leq n \leq m. \label{Appendix_eq1}
\end{equation}
%and $\max_{m \in \Omega_{\text{DD}}} 1/\gamma_{m}^* \leq \max_{m
%\in \Omega_{\text{DD}}} 1/\gamma_{m}$ for any feasible
%$\mathbf{\Gamma}=[\gamma_{1}, \gamma_{2}, ..., \gamma_{k_{\text{DD}}}]$.

We observe that the objective function in ($\text{P}_3$) is
equivalent to $\min \{\underset{1 \leq n \leq k_{\text{D}}}{\max}
\frac{1}{v_{n}}\}$ and suppose that the optimal power
allocation results in different SINRs at the first $m$ D2D links,
i.e., $\underset{1 \leq n \leq m }{\max} 1/v_{n}^*
> \underset{1 \leq n \leq m}{\min}\ 1/v_n^*$, and that the D2D
link $l_{n^*}$ has the largest SINR at first $m$ D2D links, i.e.,
$n^*=\text{arg} \ \underset{1 \leq n \leq m}{\min}\ 1/v_{n}^*$, we have
$v_{n^*}^*
> \bar v_{m}$. Besides, from (\ref{Appendix_eq1}), we
observe that $v_{n}^* $ is a strictly increasing function of
$p_{n}^*$ and is a strictly decreasing function of $p_{k}^*$ for $k
\neq n$. Therefore, there must be a small power decrement $\Delta p_{n^*}$
and SINR decrement $\Delta v_{n^*}$ for D2D link $l_{n^*}$
and a small SINR increment $\Delta v_{n}$ for other D2D
links $l_{n}$ ($ \ 1 \leq n\leq k_{\text{D}}, \ n \neq n^*$) such
that the constraints in $(\text{P}_2)$ still hold, i.e.,
\begin{equation}
\frac{(p_{n^*}^*-\Delta p)g(n^*,n^* )}{\sum_{n = 1, n \neq n^*}^{n
=k_{\text{D}} }p_{n}^*g(n, n^*)+N_{n^*}}=v_{n^*}^*-\Delta
v_{n^*} \geq \bar v_{m} \ge \bar v_{n^*},
\end{equation}
and
\begin{equation}
%\begin{split}
\frac{p_{n}^*g(n, n)}{\sum_{s =1, s \neq n^*, s \neq
s}^{k_{\text{D}}}p_{s}^*g(s, n)+(p_{n^*}^*-\Delta p_{n^*})g(n^*,
n)+N_{k}}
=v_{n}^*+\Delta v_{n} > \bar v_n.
%\end{split}
\end{equation}

Suppose that we choose a small $\Delta
v_{n}$ to satisfy $\frac{1}{(v_{n^*}^*-\Delta
v_{n^*})} \leq \underset{1 \leq n \leq m, \ n\neq n^*}{\max}\
\frac{1}{(v_{n}^*+\Delta v_{n})}$. Then, we have
\begin{equation}
\begin{split}
&\underset{1 \leq n \leq m}{\max}\left\{\frac{1}{v^*_{
n}}\right\}=\underset{1 \leq n \leq m, \ n\neq
n^*}{\max}\left\{\frac{1}{v^*_{ n}}\right\}
>\underset{1\leq n
\leq m, \ n\neq n^*}{\max}\left\{\frac{1}{v^*_{ n}+\Delta
v_{n}}\right\}\\
&=\max\left\{\underset{1 \leq n \leq m, \ n\neq
n^*}{\max}\left\{\frac{1}{v^*_{ n}+\Delta
v_{n}}\right\}, \frac{1}{v^*_{n^*}-\Delta
v_{n^*}} \right\}.
\end{split}
\end{equation}

This means that there exists another power allocation
$\mathbf{P}'\neq \mathbf{P}^* $ enabling the
SINRs at the scheduled D2D links to be $\mathbf{V}^{'} \neq \mathbf{V}^{*}$ and $\underset{1 \leq n \leq m}{\max}\ 1/\gamma_{n}^*
> \underset{1 \leq n \leq m}{\max}\ 1/v'_{n}$, where $1/v'_{n}$ is
equal to $\underset{1 \leq n \leq m, \ n\neq n^*}{\max}\ 1/(v^*_{n}+\Delta
v_{n})$ for $n \neq n^* $ and is equal to
$1/(v^*_{n^*}-\Delta v_{n^*})$ for $n = n^*$, which
causes contradiction. Thus, we have $\underset{1 \leq n \leq m}{\max}\
1/v_{n}^* = 1/v^* $, where $ \bar v_m \leq v^* \leq \bar v_{m+1}$.

Then, the optimal SINRs vector can be denoted as
$\mathbf{V}^{*}=[\underbrace {v^*, \cdots, v^*}_{m},
v_{m+1}^*, \cdots, v_{k_{\text{D}}}^*]$ and the corresponding minimum SINR is $v^*$. Next, we will prove
$v_n^*=\bar v_n$ for $ m+1 \leq n \leq k_{\text{D}}$.

Suppose that there exists a user $k^*$ satisfying
$v_{k^*}^*>\bar v_k$ for $ m+1 \leq k \leq k_{\text{D}}$.
There must be a small power decrement $\Delta p_{k^*}$ and SINR decrement
$\Delta v_{k^*}$ for D2D link $l_{k^*}$ and a small SINR
increment $\Delta v_{k}$ for D2D links $l_{k}$ ($ \ 1 \leq
k \leq m$) such that the constraints in $(\text{P}_2)$
still hold, i.e.,
\begin{equation}
\frac{(p_{k^*}^*-\Delta p_{k^*})g(k^*,k^*)}{\sum_{k = 1, k \neq k^*}^{t
=k_{\text{D}} }p_{k}^*g(k, k^*)+N_{k^*}}=v_{k^*}^*-\Delta
v_{k^*} \geq \bar v_{k^*},
\end{equation}
and
\begin{equation}
%\begin{split}
\frac{p_{k}^*g(k, k)}{\sum_{t=1, t \neq k, t \neq
k^*}^{k_{\text{D}}}p_{t}^*g(t, k)+(p_{k^*}^*-\Delta p_{k^*})g(t,
k)+N_{k}}
=v^*+\Delta v_{k}.
%\end{split}
\end{equation}

Then, if we choose a small $\Delta p_{k^*}$ and $\Delta v_{k^*}$ to satisfy $\underset{1\leq k \leq m}{\min} \ v^*+\Delta v_k \leq v_{m+1}$, the minimum SINR of the scheduled D2D links is
$\underset{1\leq k \leq m}{\min} \ v^*+\Delta v_k$, which causes contradiction. Thus,
we have $v_n^*=\bar v_n$ for $ m+1 \leq n \leq
k_{\text{D}}$.

This complete the proof of the first case.

\subsubsection{Proof of the Second Case}

The proof of the second case is similar to that of the first case
and will be omitted for page limit.

\end{document}